\documentclass[10pt,aps,twocolumn,pra,superscriptaddress]{revtex4-2}
\usepackage{braket}
\usepackage{graphicx}
\usepackage{braket}
\usepackage{amsmath}
\usepackage{dcolumn}
\usepackage[dvipsnames]{xcolor}
\usepackage{bm}
\usepackage{hyperref}
\hypersetup{
    colorlinks=true,
    linkcolor=blue,
    filecolor=blue,      
    urlcolor=blue,
    pdftitle={Overleaf Example},
    pdfpagemode=FullScreen,
    citecolor = blue
    }
\usepackage{orcidlink}

\begin{document}


\title{Pulsed single-photon spectroscopy of an emitter with vibrational coupling}
\author{Sourav Das\,\orcidlink{0000-0002-6087-846X}}
\affiliation{Department of Physics, University of Warwick, Coventry CV4 7AL, United Kingdom}

\author{Aiman Khan\,\orcidlink{0009-0006-0697-8343}}
\affiliation{Manufacturing Metrology Team, Faculty of Engineering, University of Nottingham, Nottingham NG7 2RD, United Kingdom}

\author{Elnaz Darsheshdar\,
\orcidlink{0000-0001-6341-7151}}
\affiliation{Dipartimento di Fisica, Università di Napoli Federico II, Complesso Universitario di Monte S. Angelo, Via Cintia, 80126 Napoli, Italy}

\author{Francesco Albarelli\,\orcidlink{0000-0001-5775-168X}}
\affiliation{Università di Parma, Dipartimento di Scienze Matematiche, Fisiche e Informatiche, I-43124 Parma, Italy}
\affiliation{INFN—Sezione di Milano-Bicocca, Gruppo Collegato di Parma, I-43124 Parma, Italy}

\author{Animesh Datta\,\orcidlink{0000-0003-4021-4655}}
\affiliation{Department of Physics, University of Warwick, Coventry CV4 7AL, United Kingdom}

\date{\today}

\begin{abstract}

We analytically derive the quantum state of a single-photon pulse scattered from a single quantum two-level emitter interacting with a vibrational bath. 
This solution for the quadripartite system enables an information-theoretic characterization of vibrational effects in quantum light spectroscopy. We show that vibration-induced dephasing reduces the quantum Fisher information (QFI) for estimating the emitter’s linewidth, largely reflecting the Franck–Condon suppression of light–matter coupling. Comparing time- and frequency-resolved photodetection, we find the latter to be more informative in estimating the emitter's linewidth for stronger vibrational coupling.

\end{abstract}

\maketitle

The photodynamics of a single quantum emitter is strongly governed by the interaction between its electronic and vibrational degrees of freedom~\cite{denning2019phonon}. Their effects on emission properties are probed in spectroscopies such as single emitter photoluminescence~\cite{clear2020phonon,grosso2020low}, which directly measure characteristics such as the linewidths of the zero-phonon line~(ZPL) and the second-order correlations of the emitted photons. 
Spectroscopies have employed non-classical single photon states in photoluminscent~\cite{li2023single} and coherent extinction~\cite{PhysRevLett.108.093601} methods to probe quantum light-matter interaction at the single-emitter single-photon level. 
More generally, spectroscopy using quantum light has been investigated in recent years, motivated by possible non-classical enhancements to nonlinear spectroscopy with classical light~\cite{mukamel2020roadmap}. Theoretical investigations of fundamental precision attainable in quantum-light spectroscopy has been undertaken for single-photon~\cite{albarelli2023fundamental,darsheshdar2024role,das2025optimalquantumspectroscopyusing}, and entangled biphoton~\cite{albarelli2023fundamental,khan2024does} probes. 
Information theoretic methods are also being used to optimize classical-light spectroscopy~\cite{10.1063/5.0206838}.

\begin{figure}
    \centering
    \includegraphics[width =\linewidth]{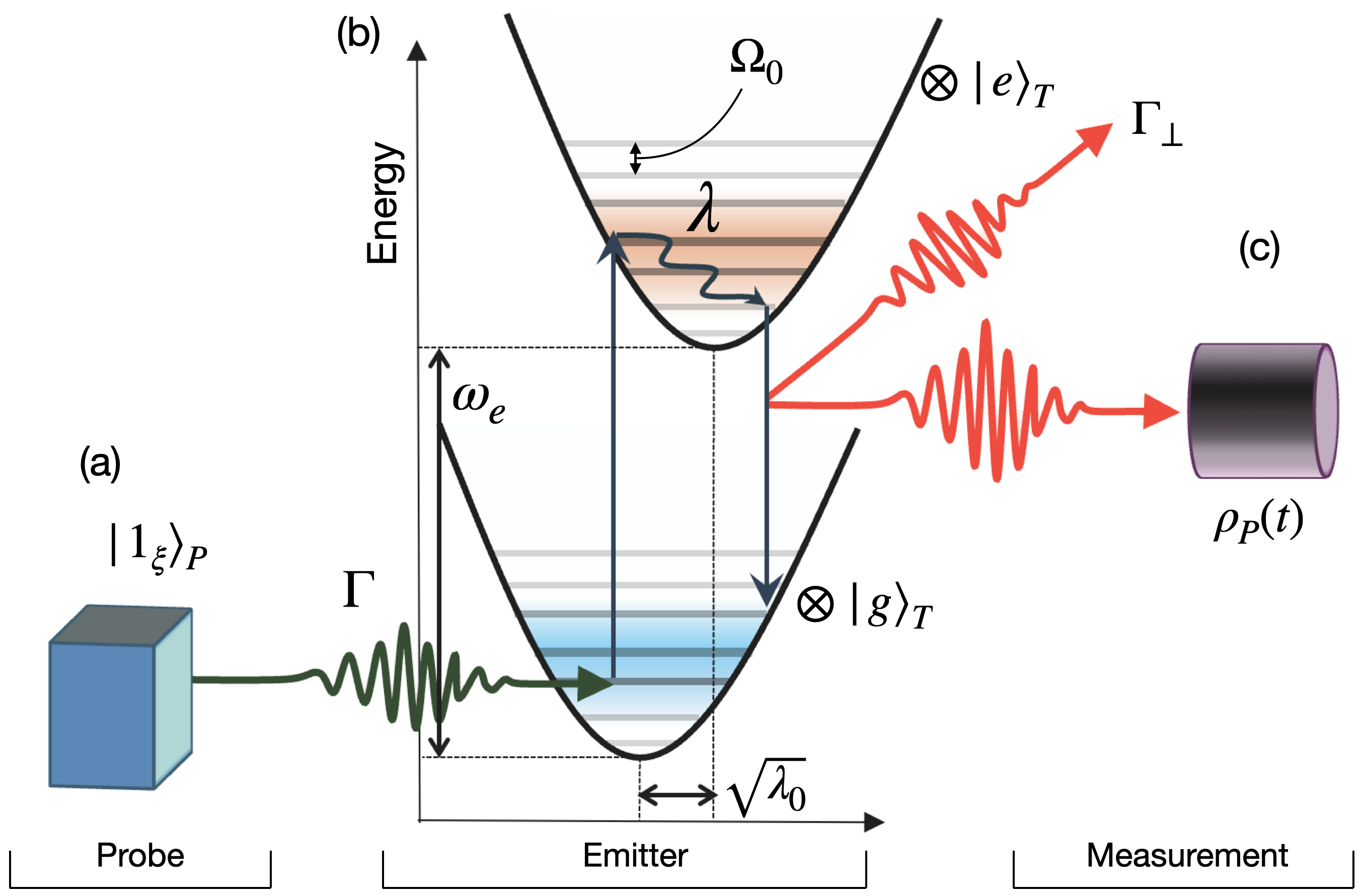}
    \caption{Single-photon scattering from a two-level emitter with vibrational coupling. A photon pulse $\ket{1_\xi}_\text{P}$ (a) excites the emitter from $\ket{g}_\text{T}$ to $\ket{e}_\text{T}$. The latter has a linewidth $\Gamma.$ Both levels feature vibrational manifolds (b). The square root of the Huang-Rhys factor $\sqrt{\lambda_0}\equiv\vert g_0\vert/\Omega_0$ sets the displacement between ground and excited vibrational manifolds. The emitter relaxes the vibrational reorganization energy $\lambda$ within the excited manifold before emitting a single photon. A fraction $\Gamma_\perp/(\Gamma_\perp + \Gamma)$ of this light is lost. The information in the detected fraction of the scattered pulse $\rho_\text{P}(t)$ is accessed via measurements (c).}
    \label{fig:1}
\end{figure}

Quantum emitters such as hexagonal boron nitride~(hBN) defects~\cite{kubanek2022coherent}, and III-V semiconductor quantum dots such as InGaAs~\cite{lodahl2015interfacing} have single, isolated optical transitions acting almost perfectly like two-level emitters (TLEs). 
In single molecules, TLEs capture the molecular-orbital transitions well~\cite{Wang2019,PhysRevResearch.2.033270,zirkelbach_high-resolution_2022,DeBernardis_2025}. The optical response of TLEs interacting with a vibrational bath has been extensively investigated using open quantum system approaches that focus on the emitter dynamics to compute absorption and emission spectra \cite{PhysRevB.87.081308,kaer2013role,PhysRevB.90.035312,PhysRevA.90.032114,PhysRevLett.120.257401,holmes2021pure,ko2022dynamics,iles2024capturing,PhysRevResearch.6.033231,PhysRevB.110.174304,Hogg_2024,PhysRevResearch.2.033270,prositto2025collisionalmodeldissipativedephasing}. 
However, the impact of the vibrational degrees of freedom on the quantum state of the scattered light and the fundamental quantum limit of single emitter spectroscopy remains completely unexplored.

A quantum information-theoretic quantification of the fundamental limits of spectroscopic precision requires characterising the quantum state of the scattered light, which encodes information about the emitter \cite{Bouchet2021,Woodworth2022,PRXQuantum.3.010354,cxvs-5pb1,rtd9-z7cl,sorelli2025ultimate}. An explicit form for the scattered radiation field is generally difficult to obtain.
It has only been possible with bipartite emitter-field systems \cite{PhysRevA.93.063807} and few tripartite systems where photonic losses are allowed for \cite{albarelli2023fundamental,darsheshdar2024role,khan2024does,khan2025tensor}.
However, none of these studies consider the vibrational degree of freedom of the emitter, which is an additional infinite-dimensional extension
 to the system composition.

In this Letter, we address the full quadripartite problem where a TLE is coupled to a set of vibrational modes that forms a displaced manifold in the excited state, illustrated in Fig.~\ref{fig:1}. 
The TLE is excited by a single-photon pulse and couples to an electromagnetic vacuum reservoir --- the former is the spectroscopic probe and the latter accounts for undetected emission (loss). We analytically solve the time-dependent Schrödinger equation in the interaction picture to obtain the absorption probability (Eq.~\eqref{excprob}) and the quantum state of the scattered pulse (Eq.~\eqref{pulset}). 
Using this state, we identify the best precision of estimating the emitter’s emission linewidth.
This precision decreases with increasing vibrational coupling due to the Franck–Condon effect (Fig.~\ref{fig:1}). We further analyze two practical single-photon measurement schemes—time-resolved and frequency-resolved photodetection—for attaining this best precision.  
We find time-resolved measurements to be more effective for weaker vibrational coupling, while frequency-resolved measurements outperform for stronger couplings. This too has a satisfying physical explanation.

Remarkably, our demonstration of Franck-Condon physics stands in contrast to previous results for mixed influence fields~\cite{PhysRevLett.123.093601, PRXQuantum.3.010321}.
These show that the same set of assumptions around the white noise approximation for light-matter coupling leads to additive dynamics at the level of master equations~(ME) and no Frank-Condon effect, if the TLE is excited using classical laser pulses. Our results challenge this conflation of ME-additivity with the lack of Franck-Condon physics through its use of single-photons states, where initial temporal correlations inherent in the photonic state produce vibrational sidebands and modulation of TLE lifetimes with increasing vibrational coupling strengths.

\textit{Model}---The TLE (T) (see Fig.~\ref{fig:1}) is characterized by its ground state ${\ket{g}}_{\text{T}},$ excited state ${\ket{e}}_{\text{T}}$,
and bare Hamiltonian $H_\text{T}=\omega_e\ket{e}_{\text{T}}\bra{e}$, $\omega_e$ being its bare transition frequency. 
 Its  coupling to the vibrational bath (V), the pulsed excitation (P), and the vacuum reservoir (E) can be compactly written (in units $\hbar=1$) by the total Hamiltonian $H(t)$ as,
\begin{equation}
\label{hamiltonian1}
\begin{aligned}
    &H(t) =  H_\text{TV} + H_\text{TPE}(t),
\end{aligned}
\end{equation}
where the constituent interactions have the form
\begin{equation}
    \label{hamiltionian2}
    \begin{aligned}
        &H_\text{TV} =  \sum_k \Omega_k c^\dagger_k c_k + \ket{e}_{\text{T}}\bra{e} \big[\lambda+
\sum_k g_k   (c_k + c_k^\dagger)\big],\\ &H_\text{TPE}(t) = i\ket{e}_{\text{T}}\bra{g} \Big(\sqrt{\Gamma}a_{\text{P}}(t) +\sqrt{\Gamma_\perp} b_{\text{E}}(t)\Big) + \text{H.c}.
    \end{aligned}
\end{equation}
Here, $c_k$ $(c_k^\dagger)$ is the vibration annihilation (creation) operator of the $k$th vibrational mode, $\Omega_k$ its frequency and $g_k$ its coupling strength to the TLE. Lastly, $\lambda = \sum_k \vert g_k\vert^2/\Omega_k$ is the reorganization energy of the vibrational bath. As $[H_\text{TV}, H_\text{T}]=0,$ Hamiltonian $H_\text{TV}$ describes a dephasing-type interaction known to be relevant to spectroscopy~\cite{skinner_pure_1986}.

The single-photon pulse and the electromagnetic vacuum (suitable for optical frequencies) reservoir couple to the TLE via a dipole interaction Hamiltonian $H_\text{TPE} = -\vec{d}\cdot(\vec{E}_\text{P}+ \vec{E}_\text{E})$ where $\vec{d}$ is the electric dipole operator of the TLE and $\vec{E}_\text{P(E)}$ is the electric field operator of the pulse (electromagnetic vacuum reservoir). 
Imposing the rotating-wave and white-noise approximations this interaction takes the form in Eq.~\eqref{hamiltionian2} (see \cite[Sec.~\ref{asec1}]{supp} for a microscopic derivation).
The so-called \emph{white-noise} operators can be expressed as Fourier transforms~(F.T.) of the frequency mode bosonic operators: $ a_\text{P}(t) = \int_{-\infty}^{\infty} d\omega \, a_\text{P}(\omega) e^{-i(\omega - \omega_c)t} / \sqrt{2\pi} $ and $ b_\text{E}(t) = \int_{-\infty}^{\infty} d\omega \, b_\text{E}(\omega) e^{-i(\omega - \omega_c)t} / \sqrt{2\pi} $ where $a_\text{P}(\omega)$ and $b_\text{E}(\omega)$ are two families of bosonic operators labeled by a continuum of frequencies.
$\omega_c$ denotes the carrier frequency of the light which is assumed to be equal to $\omega_e$ throughout. $\Gamma$ and $\Gamma_\perp$ denotes the coupling strengths corresponding to the pulse and the vacuum reservoir which also partitions the total emission rate $\Gamma+\Gamma_\perp$ into the detected and undetected parts respectively.

Due to the rotating wave approximation, $[H(t),N_\text{TPE} ]=0,$ where $N_\text{TPE}=\ket{e}_\text{T}\bra{e}+ \int_0^\infty d\omega [a_{\text{P}}^\dagger(\omega)a_{\text{P}}(\omega)+b_{\text{E}}^\dagger(\omega)b_{\text{E}}(\omega)]$ 
is the total number operator for the TLE and the electromagnetic subsystems.
This symmetry enables an analytical solution of the time-dependent Schrödinger equation following from $H(t)$ for the initial condition $\ket{\psi_\gamma(t_0)}_{\text{TVPE}} = \ket{g}_{\text{T}}\ket{\gamma}_{\text{V}}\ket{1_{\xi}}_{\text{P}}\ket{0}_{\text{E}}$
which has $N_\text{TPE}=1$. Here,
$\ket{\gamma}_{\text{V}}$ is an arbitrary pure state of the vibrational modes and $\ket{1_{\xi}}_{\text{P}}=\int_{-\infty}^{\infty}\xi(\tau)a^\dagger(\tau)\ket{0}_\text{P}$ represents the single-photon incident pulse with $\xi(\tau)$ being its temporal pulse shape~\cite{ko2022dynamics}.

Our ansatz for solving the quadripartite system is
\begin{eqnarray}
    \ket{\psi_\gamma(t)}_{\text{TVPE}} = \ket{e}_{\text{T}}\ket{A_\gamma(t)}_{\text{V}}\ket{0}_{\text{P}}\ket{0}_{\text{E}} \\ \nonumber +\ket{g}_{\text{T}} \Big(\ket{1_\gamma(t)}_{\text{PV}}\ket{0}_{\text{E}} +\ket{0}_{\text{P}}&\ket{1_\gamma(t)}_{\text{EV}} \Big),
\end{eqnarray}
where $\ket{A_\gamma(t)}_{\text{V}}$ represents the unnormalized vibrational state when the TLE is excited and $\ket{1_\gamma(t)}_{\text{PV(EV)}}$ are the unnormalised pulse-vibration (electromagnetic vacuum reservoir-vibration) joint states when the TLE is de-excited. These three unknowns are solved for in \cite[Sec.~\ref{asec2}]{supp}.

\emph{Photonic state}---We now focus on the solutions when the vibrational modes are initialized in a thermal state: $\sigma_\text{V}^{\text{th}}=\sum_\gamma \sigma_\gamma \ket{\gamma}_\text{V}\bra{\gamma}$, where for notational simplicity we assume $\ket{\gamma}_\text{V}$ to be a Fock state with number $\gamma$ and $\sigma_\gamma$ represents the corresponding thermal occupation number.
The excited state population $p_e(t) \equiv \sum_\gamma \sigma_\gamma \langle A_\gamma(t)\ket{A_\gamma(t)}_{\text{V}}$ has the analytical form 
\begin{eqnarray}
\label{excprob}
\begin{aligned}
p_e =  \Gamma \int_{t_0}^{t}\int_{t_0}^{t}dt^\prime dt^{\prime\prime}\,\xi(t^\prime)\xi^{*}(t^{\prime\prime})\,e^{{ \frac{\Gamma+\Gamma_\perp}{2}(t^\prime+t^{\prime\prime}-2t)}+\Lambda_1(t^\prime-t^{\prime\prime})}
\end{aligned}
\end{eqnarray}
where the propagator for the vibrational modes is
\begin{equation}
\label{Lmd1}
    \Lambda_1(t) =\int_0^\infty \! d\Omega\, \frac{J(\Omega)}{\Omega^2}\Big[\coth\left(\frac{\beta\Omega}{2}\right)(\cos(\Omega t)-1)
    +i\,\sin(\Omega t)\Big],
\end{equation}
 with $J(\Omega)$ being the spectral density of the TLE-vibration coupling and $\beta = (k_B T)^{-1}$ representing the inverse temperature.
 Note that we have taken the continuum limit for the vibrational modes with a spectral density function $J(\Omega)$ which takes the form $J(\Omega) = \sum_k \vert g_k\vert^2 \delta(\Omega-\Omega_k) $ for the discrete case. The propagator $\Lambda_1(t)$ for two other commonly used continuous spectral densities, the Drude-Lorentz and Brownian are presented in \cite[Sec.~\ref{asec8}]{supp}.

The quantum state of the pulse of light arriving at the detector is given by 
\begin{equation}
\label{pulset}
\begin{aligned}
    \rho_{\text{P}}(t) = \text{Tr}_{\text{TVE}}\Big(\sum_\gamma \sigma_\gamma\ket{\psi_\gamma(t)}_{\text{TVPE}}\bra{\psi_\gamma(t)}\Big)\\ = \Big(p_e(t)+p_{\Gamma_\perp}(t)\Big) \ket{0}_{\text{P}}\bra{0} + \rho^1_{\text{P}}(t),
\end{aligned}
\end{equation}
where $p_{\Gamma_\perp}= \Gamma_\perp \int_{t_0}^{\infty} d\tau p_e(\tau)$ and $\rho^1_{\text{P}}(t)$ is presented in \cite[Eq.~\eqref{eq44}]{supp}
This is our main mathematical result, and forms the basis of the rest of this work.

To explore the physics, we now focus on times long compared to the lifetime of the TLE, that is,
$t\gg 1/(\Gamma+\Gamma_\perp)$, henceforth notated as $t=+\infty.$ Then the TLE decays into the ground state $\ket{g}_\text{T}$, i.e., $p_e\rightarrow 0$ and $\ket{A_\gamma(\infty)}_{\text{V}}\rightarrow 0$. 
Thus,
\begin{eqnarray}
\begin{aligned}
\label{eqn4}
    &\rho_{\text{P}}(\infty) \equiv \sum_\gamma \sigma_\gamma \text{Tr}_\text{TVE}~\ket{\psi_\gamma(\infty)}_{\text{TVPE}}\bra{\psi_\gamma(\infty)}\noindent\\   &=p_{\Gamma_\perp}\ket{0}_{\text{P}}\bra{0}
    +\int_{t_0}^{\infty}\int_{t_0}^{\infty} d\tau d\tau^\prime\,\varrho_{\text{P}}(\tau,\tau^\prime) \ket{1_\tau}_{\text{P}}\bra{1_{\tau^\prime}},
\end{aligned}
\end{eqnarray}
where $\ket{1_\tau}\equiv a^\dagger_\text{P}(\tau)\ket{0}_\text{P}$ and $\varrho_{\text{P}}(\tau,\tau^\prime)$ is the unnormalized time-domain density matrix of the scattered pulse that arrives at the detector (see Fig (\ref{fig:1})). 
We call
\begin{eqnarray}
\begin{aligned}
\label{scatterp}
\varrho_{\text{P}}(\tau,\tau^\prime) =\xi(\tau) (\xi(\tau^\prime))^* &\\ - \Gamma\xi(\tau^\prime)^*\int_{t_0}^{\tau}dt^\prime\,\xi(t^\prime) &\,e^{{ \frac{\Gamma+\Gamma_\perp}{2}(t^\prime-\tau)}+
 \Lambda_1(t^\prime-\tau)} \\-\Gamma\xi(\tau)\int_{t_0}^{\tau^\prime}dt^\prime\,\xi(t^\prime)^* &\,e^{{ \frac{\Gamma+\Gamma_\perp}{2}(t^\prime-\tau^\prime)}+
\Lambda_1(-t^\prime+\tau^\prime)} \\
+\Gamma^2 \int_{t_0}^{\tau}\int_{t_0}^{\tau^\prime}dt^\prime dt^{\prime\prime}& \,\xi(t^\prime)\xi(t^{\prime\prime})^* \\ 
\times & e^{{ \frac{\Gamma+\Gamma_\perp}{2}(t^\prime+t^{\prime\prime}-\tau-\tau^\prime)}
\,+\Lambda_2(t^\prime,t^{\prime\prime},\tau,\tau^\prime)},
\end{aligned}
\end{eqnarray}
the temporal density matrix (TDM) of the scattered pulse, and 
\begin{eqnarray}
\label{lm2}
\Lambda_2(t^\prime,t^{\prime\prime},\tau,\tau^\prime) = \Lambda_1(t^\prime-\tau)+\Lambda_1(t^{\prime\prime}-\tau^\prime)^*-&\Lambda_1(t^\prime-\tau^\prime)\nonumber\\ - \Lambda_1(t^{\prime\prime}-\tau)^* + \Lambda_1(t^\prime-t^{\prime\prime}) + \Lambda_1(\tau-\tau^\prime).
\end{eqnarray}
This TDM forms a mixed state within the single-excitation subspace of the pulse, due to entanglement with the vibrational environment captured by the last four terms on the right hand side of Eq.~\eqref{lm2}. These survive even when the vibrational modes are at zero temperature.

\textit{Spectroscopy}---
Spectroscopy typically comprises of inferring 
the parameters that model the emitter such as $\Gamma,$ $\omega_e$ and $\lambda_0$.
The precision of any unbiased estimator $\hat{\theta}$ for parameter $\theta$ satisfies 
\begin{eqnarray}
    \text{Var}(\hat{\theta})\geq \frac{1}{\mathcal{F}(\rho_{\text{P}}(\infty),\{\Pi_x\})}\geq\frac{1}{\mathcal{Q}(\rho_{\text{P}}(\infty))}.
\end{eqnarray}
The second inequality is known as the quantum Cr\'{a}mer-Rao bound \cite{helstrom1967minimum,PhysRevLett.72.3439}.
It is asymptotically attainable for individual parameters. The quantum Fisher information (QFI) $\mathcal{Q}(\rho_{\text{P}}(\infty))$ quantifies the best precision with which 
$\theta$ can be estimated from the quantum state $\rho_{\text{P}}(\infty)$.
The classical Fisher information (CFI) $\mathcal{F}(\rho_{\text{P}}(\infty),\{\Pi_x\})$ is the amount of information recovered via the measurement characterised by the positive-operator-valued measure~(POVM) $\{\Pi_x\}$. 

\begin{figure}
    \centering
    \includegraphics[width=1\linewidth]{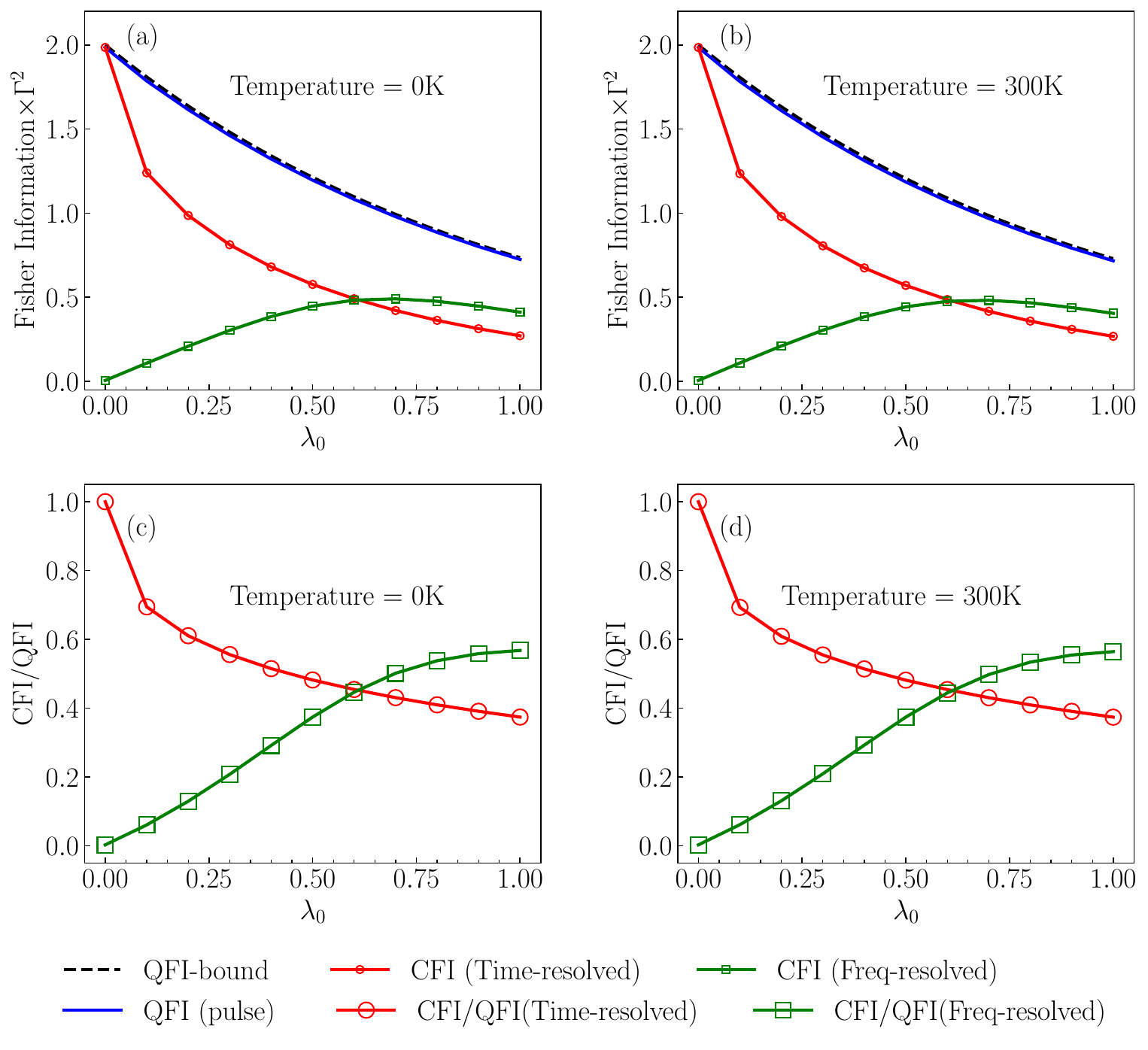}
    \caption{Fisher information (in the units of $\Gamma^2$) of estimating $\Gamma$ from the scattered pulse as a function of the Huang-Rhys factor $\lambda_0$.  The curves display the QFI upper-bound $\mathcal{Q}_{\text{bound}}$ (black dashed), QFI of the scattered pulse $\mathcal{Q}(\rho_{\text{P}}(\infty))$ (blue solid) and CFI for time- (red small circles) and frequency-resolved (green small squares) photon counting. (c,d) plots the corresponding ratios of the CFI and QFI of the scattered pulse. The incident pulse-shape is a decaying exponential (in the time-domain) $\xi(t)= \exp(-t/(2T_\sigma))\Theta(t)/\sqrt{T_\sigma}$ with pulse-duration $T_\sigma=1/\Gamma$. We evaluate this state numerically via the convolution theorem \cite[Sec.~\ref{asec3}]{supp}. The system parameters are: $\Gamma=0.15\text{THz}$, $\Omega_0=1000\text{cm}^{-1}$.}
\label{fig:qfi}
\end{figure}

Here, we will specialise our discussion to the estimation of linewidth $\Gamma$ of the TLE emission, and investigate the effect of the vibrational coupling on the Fisher information of the scattered pulse. We assume a perfect emitter-pulse coupling, i.e., $\Gamma_\perp=0$ for simplicity (See \cite[Sec.~\ref{asec_imperfect}]{supp} for results on $\Gamma_\perp \neq 0.$).
In this scenario, the TDM $\varrho_{\text{P}}(\tau,\tau^\prime)$ carries the full description of the scattered pulse and is a normalized density matrix according to Eq.~\eqref{eqn4}. 
For conceptual clarity, we also assume the vibration to be be comprised of a single vibrational mode. This accurately represents structured environments coupling to single-molecule emitters \cite{ADOLPHS20062778} for which the vibrational spectral density $J(\Omega)=\lambda_0\Omega_0^2\delta(\Omega-\Omega_0)$, where $\lambda_0=\vert g_0\vert^2/\Omega_0^2$ is the Huang-Rhys factor which is a dimensionless number representing the vibrational coupling strength and $\Omega_0$ is the vibration frequency. Results for estimating $\Gamma$ when the TLE is coupled to a continuum of vibrational modes such as Drude-Lorentz and Brownian are presented in \cite[Sec.~\ref{asec9}]{supp}.

The QFI of the scattered pulse $\mathcal{Q}(\rho_{\text{P}}(\infty))$ is dependent on the state itself and its parametric derivative $\partial_\Gamma\rho_{\text{P}}(\infty)$. In the time domain, we evaluate $\varrho_{\text{P}}(\tau,\tau^\prime)$ and $\partial_\Gamma\varrho_{\text{P}}(\tau,\tau^\prime)$ numerically on a $N\times N$ grid of equally spaced time-bins: $(\tau,\tau^\prime)\in [\tau_0, \tau_{N-1}]\times[\tau_0, \tau_{N-1}]$ \cite[Sec.~\ref{asec3}]{supp}. The computational cost scales as $\mathcal{O}(N^3 \text{log}(N))$. 
To capture all the timescales of the system, we choose $0\leq\lambda_0\leq 1$, $\tau_0=-10/\Gamma$, $\tau_{N-1}=10/\Gamma$, and $N=10000$. 
The QFI $\mathcal{Q}(\rho_{\text{P}}(\infty))= \text{Tr}(\rho_{\text{P}}(\infty)L_\Gamma^2)$ is calculated by numerically solving the Lyapunov equation $L_\Gamma\rho_{\text{P}}(\infty)+\rho_{\text{P}}(\infty)L_\Gamma=2\partial_\Gamma\rho_{\text{P}}(\infty)$ for the symmetric logarithmic derivative (SLD) operator $L_\Gamma.$

Fig.~\ref{fig:qfi}(a,b) shows that the QFI for estimating $\Gamma$ from the scattered pulse decreases as $\lambda_0$ increases due to the Franck--Condon effect: the reduced overlap between the vibrational wavefunctions of the ground and excited manifolds weakens the effective light–matter coupling \cite{PhysRevLett.94.206804,PhysRevB.74.205438,leturcq2009franck}. 

To quantify this effect, we calculate an analytical upper-bound on the QFI $\mathcal{Q}(\rho_\text{P}(\infty)) \leq \mathcal{Q}_{\text{bound}}$ that keeps track of the entangled quantum state of the vibrational and pulse mode. As shown in \cite[Sec.~\ref{asec6}]{supp},
\begin{eqnarray}
\label{qfibound}
    \mathcal{Q}_{\text{bound}} = \sum_{k=-\infty}^{\infty} f_k \int_{-\infty}^{\infty} \! d\omega \, g(\omega-k\Omega_0) \vert\tilde{\xi}(\omega)\vert^2,
\end{eqnarray}
 $g(\omega)=64\omega^2/(\Gamma^2+4\omega^2)^2$, $\tilde{\xi}(\omega)=\int_{-\infty}^{\infty}d\tau e^{-i\omega\tau}\xi(\tau)/\sqrt{2\pi}$ is the incident pulse in the frequency domain, and 
 \begin{eqnarray}
\label{frank}
    f_k = e^{-\lambda_0  \bar{n}} I_k\left(\lambda_0 \sqrt{\bar{n}^2-1}\right)\left(\frac{\bar{n}+1}{\bar{n}-1}\right)^{k/2} ,
\end{eqnarray}
where $I_k(x)$ is the modified Bessel function of order $k$, and $\bar{n}=\coth(\beta\Omega_0/2)$.
Notably, the $f_k$ are known in the theory of light absorption from the work of Huang and Rhys~\cite[Eq.~(5.1)]{huang1950theory}.
 In our formulation, they emerge from the Jacobi–Anger expansion \cite[Sec.~\ref{asec4}]{supp} of $e^{\Lambda_1(t)}=\sum_{k=-\infty}^\infty f_k e^{i k\Omega_0}.$
 At zero temperature ($\bar{n}=1$), the coefficients $f_k$ reduce to the Franck–Condon factors $\tilde{f}_k \equiv \vert \langle k \vert \mathcal{D}(\sqrt{\lambda_0}) \vert 0 \rangle_\text{V} \vert^2 = e^{-\lambda_0}(\lambda_0)^k/k!$ \cite{PhysRevA.109.042817} which quantify the overlap between the vibrational configurations of the displaced vibronic manifolds (see Fig.~\ref{fig:1}), where $\mathcal{D}(x)=e^{x c-x^*c^\dagger}$ is the vibrational displacement operator. We refer to the $f_k$ as modified Franck–Condon factors.

Eq.~\eqref{qfibound} can be interpreted as the information associated with each vibrational transition, weighted by the modified Franck-Condon factors $f_k$, while the dependence on $k\Omega_0$ is due to the detuning induced by the vibrational transition from the ground state to the $k$th excited state.
In the limit $\Gamma\ll\Omega_0$ and $T_\sigma\Omega_0\gg 
1$ \footnote{These limits should be read as additional constraints on the system beyond the standard weak-coupling and the narrow-band limits: $\Gamma\ll\omega_e$ and $T_\sigma\omega_e\gg 
1$ imposed on the emitter-light interaction, where $\omega_e$ is the bare transition frequency of the emitter.} with $T_\sigma$ being the pulse duration defined as the standard deviation of $\vert\xi(\tau)\vert^2$, the contribution of the side-bands in $\mathcal{Q}_\text{bound}$ ($k\neq0$ in Eq.~\eqref{qfibound}) is negligible compared to the contribution of the ZPL ($k=0$):
\begin{equation}
\label{zpl}
    \mathcal{Q}_{\text{bound}} \approx f_0 \mathcal{Q}_{\text{no-vibration}},
\end{equation}
where $\mathcal{Q}_{\text{no-vibration}}$ is the QFI of the pulse in the absence of vibration ($\lambda_0=0$), which is expressed by $\mathcal{Q}_{\text{no-vibration}}=\int_{-\infty}^{\infty}d\omega g(\omega)\vert\tilde{\xi}(\omega)\vert^2$ \cite{darsheshdar2024role,das2025optimalquantumspectroscopyusing}. Eq.~\eqref{zpl} shows that the reduction of the QFI with the vibrational coupling is directly related to the decrease of the modified Franck-Condon factors with increasing $\lambda_0$. At zero temperature, $f_0 = e^{-\lambda_0}$ which indicates that $\mathcal{Q}_{\text{bound}}$  exponentially decreases with $\lambda_0$ in Fig.~\ref{fig:qfi}(a). 

The QFI of the pulse is close (but not identical~\footnote{At zero temperature, this upper-bound corresponds to the QFI of the pulse-vibration pure state neglecting some negative contributions. See \cite[Sec.~\ref{asec6}]{supp}.}) to $\mathcal{Q}_{\text{bound}}$ in Fig.~\ref{fig:qfi}(a,b), because the modified Franck–Condon factors $f_k$ decay rapidly for higher-order vibrational transitions $k.$ So only a few vibrational levels are significantly populated during excitation. As a result, the scattered pulse and vibrations remain only weakly entangled. At finite temperatures, the QFI bound employs the extended convexity property~\cite{PhysRevA.91.042104,PhysRevA.93.042121}, which makes it less tight (see \cite[Sec.~\ref{asec6}]{supp}). However, for $\Omega_0=1000\text{cm}^{-1}$, at room temperature $\bar{n}\approx 1.02,$ close to that for zero temperature ($\bar{n}= 1$). Therefore, the corresponding curves in Fig.~\ref{fig:qfi}(a,b) are close. For a more exaggerated temperature dependence, see \cite[Fig.~\ref{fig:qfi100}]{supp}.

We now discuss the CFI attainable from two common single-photon measurements: time-resolved photodetection~(TRP) with the POVM $\{\ket{1_\tau}_\text{P}\bra{1_\tau}\}$ and frequency-resolved photodetection~(FRP) with the POVM $\{\ket{1_\omega}_\text{P}\bra{1_\omega}\}$, where $\ket{1_\omega}_\text{P}=a^\dagger_\text{P}(\omega)\ket{0}_\text{P}$ \cite{PhysRevLett.123.133602,PhysRevLett.121.083602,Duan:24}. The former measures the temporal profile of the detected light, corresponding to the diagonal elements of the TDM in Eq.~\eqref{scatterp}.
The latter measures the spectral profile of the scattered pulse—that is, the diagonal elements of the F.T. of the TDM (see \cite[Sec.~\ref{asec5}]{supp}). These elements are directly accessed from the numerically evaluated $\rho_\text{P}(\infty)$ which is then used to compute the corresponding CFI.

Fig.~\ref{fig:qfi}(c,d) compares the ratio of the CFI obtained from these two measurements to the QFI for different vibrational coupling strengths. At $\lambda_0 = 0$, the TRP fully recovers the QFI, consistent with previous results~\cite{albarelli2023fundamental,darsheshdar2024role}, whereas the FRP is completely insensitive to $\Gamma$. 
This is because the scattering is completely elastic, leaving the frequency profile of the scattered pulse unperturbed.
For nonzero vibrational coupling ($\lambda_0 > 0$), Raman transitions modify this frequency profile and generate vibrational sidebands. These sidebands are detuned, as discussed earlier, which render the TRP suboptimal in recovering the QFI. Consequently, the CFI/QFI ratio for the TRP decreases with $\lambda_0$ [Fig.~\ref{fig:qfi}(c,d)]. In contrast, because the linewidths of the vibrational sidebands scale with $\Gamma$, the FRP becomes increasingly sensitive to $\Gamma$, leading to an increase in its CFI/QFI ratio [Fig.~\ref{fig:qfi}(c,d)]. Physically, the information from the TLE is redistributed into the vibrational modes of the emitter; hence, full recovery of the QFI requires simultaneous measurement of both the amplitudes and phases of the TDM of the scattered pulse.  

At {stronger} vibrational coupling ($\lambda_0 \gtrsim 0.5$) \cite{PhysRevA.109.042817}, the FRP recovers more QFI than the TRP. 
This is because time-resolved measurements see only the populations in the diagonals of TDM but none of the phase fluctuations due to $\Lambda_1$ and $\Lambda_2$ in Eq.~\eqref{scatterp}. In contrast, the frequency-resolved measurements capture the redistribution of emission across vibration-mediated channels, preserving more information about the linewidth. This effect is also observed for estimation with imperfect detectors ($\Gamma_\perp>0$) presented in \cite[Fig.~\ref{fig:lossy}]{supp} and for continuum vibrational modes \cite[Fig.~\ref{qfi-cont}]{supp}.

The quantum state of the light pulse in Eq.~\eqref{pulset} can also be used to evaluate the precision of estimating the frequencies $\omega_e,\Omega_0$ and temperature $\beta$~\cite{Meng2024,Kocheril2025,PRXQuantum.4.040314}. 
We present results on the estimation of the Huang-Rhys factor~\cite{Burt1981,PhysRevA.109.042817,Gelin2020,Whalley2021} in \cite[Sec.~\ref{asec7}]{supp}).

\textit{Conclusion}---We have studied the scenario of a single-photon pulse scattered by a two-level emitter coupled to vibrational modes. Obtaining an analytical expression for the quantum state of scattered pulse, we identified the information-theoretic limits to the precision of single-photon, single-emitter photoluminescence spectroscopy. We find that the fundamental spectroscopic precision diminishes with increasing vibrational coupling, a consequence of the Franck-Condon effect.

Future studies should seek practical spectroscopic measurements independent of the unknown emitter parameters whose CFI attains the QFI, as well as optimal pulse shapes~\cite{das2025optimalquantumspectroscopyusing}. They could also go beyond the white-noise approximation for a deeper understanding of the non-Markovianity induced by non-stationary electromagnetic bath in explicitly non-additive formulations, and their effect on single-emitter spectroscopic precision. This is expected to be crucial for cavity-enhanced field-emitter couplings, where the increased strength of light-matter couplings leads to breakdown of additive MEs, and collective coordinate transformations~\cite{PhysRevLett.123.093601}, or tensor network methods must be applied. While these non-additive approaches have so far examined photon correlation functions, characterising the limits to spectroscopic precision requires the evaluation of entropic quantities like the QFI, a significantly harder endeavor. Our work could perhaps light a path.

\textit{Acknowledgments---}We thank Benjamin Brecht, Jake Iles-Smith, Denis Seletskiy, and Andrea Smirne for fruitful discussions. This work has been funded, in part, by an EPSRC New Horizons grant (EP/V04818X/1) and the UKRI (Reference Number: 10038209) under the UK Government’s Horizon Europe Guarantee for the Research and Innovation Programme under agreement 101070700 (MIRAQLS). Computing facilities were provided by the Scientific Computing Research Technology Platform of the University of Warwick.

\let\oldaddcontentsline\addcontentsline
\renewcommand{\addcontentsline}[3]{}
\bibliographystyle{apsrev4-2}
\bibliography{bib}
\let\addcontentsline\oldaddcontentsline


\clearpage

\onecolumngrid
\renewcommand{\theequation}{S.\arabic{equation}}
\setcounter{equation}{0}

\begin{center}
    \textbf{\large{Supplementary materials}}
\end{center}

\tableofcontents

\section{Microscopic derivation of the light-matter Hamiltonian}
\label{asec1}

In this section, we derive the total light-matter Hamiltonian in Eq.~\eqref{hamiltionian2} from the dipole interaction Hamiltonian $H_\text{TPE} = -\vec{d}\cdot(\vec{E}_\text{P}+ \vec{E}_\text{E})$. The dipole operator of the emitter takes the following form:
\begin{eqnarray}
    \vec{d} =  \vec{d}_{ge} \ket{e}_\text{T}\bra{g} \,+\, \text{h.c.} ,
\end{eqnarray}
where $ \vec{d}_{ge}=\bra{g}\vec{d}\ket{e}_\text{T}$ is the relevant matrix elements of the dipole operator. 

The linearly polarised electric field operator in the paraxial regime is given by:
\begin{equation}
\label{eq:E}
    \vec{E}_\text{P} = i \int_{0}^{\infty} d\omega \,\vec{\mathcal{E}}_\text{P}(\omega)\,a_\text{P}(\omega)  + \text{h.c.}, \text{ and, }\vec{E}_\text{E} = i \int_{0}^{\infty} d\omega \,\vec{\mathcal{E}}_\text{E}(\omega)\,b_\text{E}(\omega)  + \text{h.c.}
\end{equation}
where $\vec{\mathcal{E}}_\text{P(E)}(\omega)$ represents the amplitude of the electric field corresponding to the Pulse (vacuum reservoir) mode.

Transforming to the interaction picture with respect to the static Hamiltonian $$H_0 = H_\text{T}+H_\text{P}+H_\text{E} = \omega_e \ket{e}_\text{T}\bra{e}+ \int_{0}^{\infty} d\omega \,\omega a_\text{P}^\dagger(\omega) a_\text{P}(\omega)+ \int_{0}^{\infty} d\omega \,\omega b_\text{E}^\dagger(\omega) b_\text{E}(\omega)$$ and rotating away the rapidly
oscillating terms, the total light-matter Hamiltonian takes the following form
\begin{eqnarray}
    H_{\text{TPE}}(t) =  
    i \ket{e}_\text{T}\bra{g} \int_{0}^{\infty} d\omega \Big[ \vec{d}_{ge}\cdot\vec{\mathcal{E}}_\text{P}(\omega)\,a_\text{P}(\omega)+\vec{d}_{ge}\cdot\vec{\mathcal{E}}_\text{E}(\omega)\,b_\text{E}(\omega) \Big]e^{-i(\omega-\omega_e)t} +\text{h.c}
\end{eqnarray}
We now assume that the field amplitudes has a finite bandwidth $2B$ around a carrier frequency $\omega_c$ such that $\vert\vec{\mathcal{E}}_\text{P/E}(\vert\omega-\omega_c\vert>B)\vert\approx 0$. With this assumption, the light-matter Hamiltonian can be rewritten as: 

\begin{eqnarray}
    H_{\text{TPE}}(t) =  
    i \ket{e}_\text{T}\bra{g} \int_{\omega_c-B}^{\omega_c+B} d\omega \Big[ \vec{d}_{ge}\cdot\vec{\mathcal{E}}_\text{P}(\omega)\,a_\text{P}(\omega)+\vec{d}_{ge}\cdot\vec{\mathcal{E}}_\text{E}(\omega)\,b_\text{E}(\omega) \Big]e^{-i(\omega-\omega_e)t} +\text{h.c}.
\end{eqnarray}

Now, we impose the white-noise approximation: $B<<\omega_c$. Consequentially, the amplitude of the field operators become only dependent on $\omega_c$ and the limits to the integral over the frequency modes can be set to $-\infty$ to $+\infty$. Thus, we can write,
\begin{eqnarray}
    H_{\text{TPE}}(t) =  i\ket{e}_\text{T}\bra{g}\left( \sqrt{\Gamma}  a_\text{P}(t) + \sqrt{\Gamma_\perp}  b_\text{E}(t) \right) +\text{h.c},
\end{eqnarray}
where the coupling strengths $$\Gamma = \left(\vec{d}_{ge}\cdot\vec{\mathcal{E}}_\text{P}(\omega_c)\right)^2, \text{ and, }\Gamma_\perp = \left(\vec{d}_{ge}\cdot\vec{\mathcal{E}}_\text{E}(\omega_c)\right)^2.$$

\section{Derivation of the final quantum state of the scattered pulse}
\label{asec2}
For clarity, we first introduce the following notations to represent the wave-packets of the pulse and the vacuum reservoir: $\ket{1_\gamma(t)}_{\text{PV}}\equiv \int d\tau \,\ket{\chi^\gamma_{\text{P}}(t,\tau)}_{\text{V}} a^\dagger_{\text{P}}(\tau)\ket{0}_\text{P}$ and $\ket{1_\gamma(t)}_{\text{EV}}\equiv \int d\tau \,\ket{\chi^\gamma_{\text{E}}(t,\tau)}_{\text{V}} b^\dagger_{\text{E}}(\tau)\ket{0}_\text{E}$ where, $\ket{\chi^\gamma_{\text{P}}(t,\tau)}_{\text{V}}$ and $\ket{\chi^\gamma_{\text{E}}(t,\tau)}_{\text{V}}$ are vibrational states.
With this notation, the ansatz reads,

\begin{equation}
    \ket{\psi_\gamma(t)}_{\text{TVPE}} = \ket{e}_{\text{T}}\ket{A_\gamma(t)}_{\text{V}}\ket{0}_{\text{P}}\ket{0}_{\text{E}} + \ket{g}_{\text{T}}\Bigg(\int d\tau \,\ket{\chi^\gamma_{\text{P}}(t,\tau)}_{\text{V}} a^\dagger_{\text{P}}(\tau)+\int d\tau \,\ket{\chi^\gamma_{\text{E}}(t,\tau)}_{\text{V}} b^\dagger_{\text{E}}(\tau)\Bigg)\ket{0}_{\text{P}}\ket{0}_{\text{E}}.
\end{equation}
This ansatz will be used to solve the Schrodinger equation for $H(t)$,
\begin{equation}
    \frac{\partial}{\partial t}\ket{\psi_\gamma(t)}_{\text{TVPE}} = -i H(t)\ket{\psi_\gamma(t)}_{\text{TVPE}}.
\end{equation}

The action of the Hamiltonian on the ansatz obtains the following,
\begin{equation}
\begin{split}
    i H(t)\ket{\psi_\gamma(t)}_{\text{TVPE}} &= i H_{\text{TV}}\ket{\psi_\gamma(t)}_{\text{TVPE}} 
    \\&-\sqrt{\Gamma} a^\dagger_{\text{P}}(t)\ket{g}_{\text{T}}\ket{A_\gamma(t)}_{\text{V}}\ket{0}_{\text{P}}\ket{0}_{\text{E}} - \sqrt{\Gamma_\perp} b^\dagger_{\text{E}}(t)\ket{g}_{\text{T}}\ket{A_\gamma(t)}_{\text{V}}\ket{0}_{\text{P}}\ket{0}_{\text{E}} \\
     +\ket{e}_{\text{T}}\Bigg(\int d\tau &\sqrt{\Gamma}\,\ket{\chi^\gamma_{\text{P}}(t,\tau)}_{\text{V}} a_{\text{P}}(t)a^\dagger_{\text{P}}(\tau)+\int d\tau \,\sqrt{\Gamma_\perp}\ket{\chi^\gamma_{\text{E}}(t,\tau)}_{\text{V}} b_{\text{E}}(t)b^\dagger_{\text{E}}(\tau)\Bigg)\ket{0}_{\text{P}}\ket{0}_{\text{E}} .
\end{split}
\end{equation}
After simplifications,
\begin{equation}
\begin{split}
    i H(t)\ket{\psi_\gamma(t)}_{\text{TVPE}} &= i H_{\text{TV}}\ket{\psi_\gamma(t)}_{\text{TVPE}} 
    \\-\sqrt{\Gamma}\int d\tau \,\delta(t-\tau) & a^\dagger_{\text{P}}(\tau)\ket{g}_{\text{T}}\ket{A_\gamma(\tau)}_{\text{V}}\ket{0}_{\text{P}}\ket{0}_{\text{E}} - \sqrt{\Gamma_\perp} \int d\tau \,\delta(t-\tau)b^\dagger_{\text{E}}(\tau)\ket{g}_{\text{T}}\ket{A_\gamma(\tau)}_{\text{V}}\ket{0}_{\text{P}}\ket{0}_{\text{E}} \\&
     +\ket{e}_{\text{T}}\Bigg(\sqrt{\Gamma}\,\ket{\chi^\gamma_{\text{P}}(t,t)}_{\text{V}} +\sqrt{\Gamma_\perp}\,\ket{\chi^\gamma_{\text{E}}(t,t)}_{\text{V}}\Bigg)\ket{0}_{\text{P}}\ket{0}_{\text{E}} .
\end{split}
\end{equation}

Differentiating the ansatz with respect to time we get,
\begin{equation}
\begin{split}
    \frac{\partial}{\partial t}\ket{\psi_\gamma(t)}_{\text{TVPE}} &= \ket{e}_{\text{T}}\frac{\partial}{\partial t}\ket{A_\gamma(t)}_{\text{V}}\ket{0}_{\text{P}}\ket{0}_{\text{E}}  \\& +\ket{g}_{\text{T}}\Bigg(\int d\tau \,\frac{\partial}{\partial t}\ket{\chi^\gamma_{\text{P}}(t,\tau)}_{\text{V}} a^\dagger_{\text{P}}(\tau)+\int d\tau \,\frac{\partial}{\partial t}\ket{\chi^\gamma_{\text{E}}(t,\tau)}_{\text{V}} b^\dagger_{\text{E}}(\tau)\Bigg)\ket{0}_{\text{P}}\ket{0}_{\text{E}}.
\end{split}
\end{equation}

Putting it together in the Schrodinger equation we get the following set of coupled differential equations,

\begin{equation}
\label{eq17}
\begin{split}
    \frac{\partial}{\partial t}\ket{A_\gamma(t)}_{\text{V}}\ket{e}_{\text{T}} = -i H_{\text{TV}} \ket{A_\gamma(t)}_{\text{V}}\ket{e}_{\text{T}} +\sqrt{\Gamma}\,\ket{\chi^\gamma_{\text{P}}(t,t)}_{\text{V}}\ket{e}_{\text{T}} +\sqrt{\Gamma_\perp}\,\ket{\chi^\gamma_{\text{E}}(t,t)}_{\text{V}}\ket{e}_{\text{T}},
\end{split}
\end{equation}

\begin{equation}
\label{eq18}
\begin{split}
    \frac{\partial}{\partial t}\ket{\chi^\gamma_{\text{P}}(t,\tau)}_{\text{V}}\ket{g}_{\text{T}} = -i H_{\text{TV}} \ket{\chi^\gamma_{\text{P}}(t,\tau)}_{\text{V}}\ket{g}_{\text{T}} -\sqrt{\Gamma}\,\delta(t-\tau)\ket{A_\gamma(\tau)}_{\text{V}}\ket{g}_{\text{T}} ,
\end{split}
\end{equation}
and,
\begin{equation}
\label{eq19}
\begin{split}
    \frac{\partial}{\partial t}\ket{\chi^\gamma_{\text{E}}(t,\tau)}_{\text{V}}\ket{g}_{\text{T}} = -i H_{\text{TV}} \ket{\chi^\gamma_{\text{E}}(t,\tau)}_{\text{V}}\ket{g}_{\text{T}} -\sqrt{\Gamma_\perp}\,\delta(t-\tau)\ket{A_\gamma(\tau)}_{\text{V}}\ket{g}_{\text{T}} .
\end{split}
\end{equation}

We can integrate Eq.~\eqref{eq18}, and Eq.~\eqref{eq19} using the integrating factor $e^{i t H_{\text{TV}}}$ as follows,
\begin{equation}
\begin{split}
    \ket{\chi^\gamma_{\text{P}}(t,\tau)}_{\text{V}}\ket{g}_{\text{T}} = \xi(\tau)e^{-i H_{\text{TV}} t}\ket{g}_{\text{T}}\ket{\gamma}_{\text{V}} -\sqrt{\Gamma} \int_{t_0}^{t} dt^\prime \,\delta(t^\prime-\tau)e^{-i H_{\text{TV}} (t^\prime-t)}\ket{A_\gamma(\tau)}_{\text{V}}\ket{g}_{\text{T}},
\end{split}
\end{equation}

\begin{equation}
\begin{split}
    \ket{\chi^\gamma_{\text{E}}(t,\tau)}_{\text{V}}\ket{g}_{\text{T}} =  -\sqrt{\Gamma_\perp} \int_{t_0}^{t} dt^\prime \,\delta(t^\prime-\tau)e^{-i H_{\text{TV}} (t^\prime-t)}\ket{A_\gamma(\tau)}_{\text{V}}\ket{g}_{\text{T}},
\end{split}
\end{equation}
After simplifications, we obtain
\begin{equation}
\label{eq22}
\begin{split}
    \ket{\chi^\gamma_{\text{P}}(t,\tau)}_{\text{V}} = \xi(\tau)e^{-i H^g_{\text{V}} t}\ket{\gamma}_{\text{V}} -\sqrt{\Gamma}  \,\Theta(t-\tau)e^{-i H^g_{\text{V}} (t-\tau)}\ket{A_\gamma(\tau)}_{\text{V}},
\end{split}
\end{equation}
and
\begin{equation}
\label{eq23}
\begin{split}
    \ket{\chi^\gamma_{\text{E}}(t,\tau)}_{\text{V}} =  -\sqrt{\Gamma_\perp} \,\Theta(t-\tau)e^{-i H^g_{\text{V}} (t-\tau)}\ket{A_\gamma(\tau)}_{\text{V}},
\end{split}
\end{equation}
where $H^g_{\text{V}}= \bra{e}H_{\text{TV}}\ket{e}_\text{T}$. Putting Eq.~\eqref{eq22}, and Eq.~\eqref{eq23} in Eq.~\eqref{eq17} we obtain,
\begin{equation}
\label{eq24}
\begin{split}
    \frac{\partial}{\partial t}\ket{A_\gamma(t)}_{\text{V}}\ket{e}_{\text{T}} = -i H_{\text{TV}} \ket{e}_{\text{T}}\ket{A_\gamma(t)}_{\text{V}} + \sqrt{\Gamma}\xi(t)e^{-i H^g_{\text{V}} t}\ket{e}_{\text{T}}\ket{\gamma}_{\text{V}} - \frac{\Gamma+\Gamma_\perp}{2}\ket{e}_{\text{T}}\ket{A_\gamma(t)}_{\text{V}}
\end{split}
\end{equation}
Using the integrating factor $\exp\Big({i\Big( H_{\text{TV}} + \frac{\Gamma+\Gamma_\perp}{2}\ket{e}_{\text{T}}\bra{e}\Big)t}\Big)$ we can integrate Eq.~\eqref{eq24} as follows,

\begin{equation}
\label{eq25}
\begin{split}
    \ket{A_\gamma(t)}_{\text{V}}\ket{e}_{\text{T}} = \sqrt{\Gamma}\int_{t_0}^{t}dt^\prime\,\xi(t^\prime) \,\exp\Big({i\Big( H_{\text{TV}} + \frac{\Gamma+\Gamma_\perp}{2}\Big)\ket{e}_{\text{T}}\bra{e}(t^\prime-t)}\Big)e^{-i H^g_{\text{V}} t^\prime}\ket{e}_{\text{T}}\ket{\gamma}_{\text{V}}
\end{split}
\end{equation}

After simplification,
\begin{equation}
\label{eq26}
\begin{split}
\ket{A_\gamma(t)}_{\text{V}} = \sqrt{\Gamma}\int_{t_0}^{t}dt^\prime\,\xi(t^\prime) \,\exp\Big({i\Big( H^e_{\text{V}} + \frac{\Gamma+\Gamma_\perp}{2}\Big)(t^\prime-t)}\Big)e^{-i H^g_{\text{V}} t^\prime}\ket{\gamma}_{\text{V}}
\end{split}
\end{equation}
where $H^e_{\text{V}} = \bra{e} H_{\text{TV}}\ket{e}_{\text{T}}$
Putting this solution in Eq.~\eqref{eq22} and Eq.~\eqref{eq23}, we obtain,
\begin{equation}
\label{eq27}
\ket{\chi^\gamma_{\text{P}}(t,\tau)}_{\text{V}} = \xi(\tau)e^{-i H^g_{\text{V}} t}\ket{\gamma}_{\text{V}} -\Gamma  \,\Theta(t-\tau)\int_{t_0}^{\tau}dt^\prime\,\xi(t^\prime) \,e^{-i H^g_{\text{V}} (t-\tau)}\exp\Big({i\Big( H^e_{\text{V}} + \frac{\Gamma+\Gamma_\perp}{2}\Big)(t^\prime-\tau)}\Big)e^{-i H^g_{\text{V}} t^\prime}\ket{\gamma}_{\text{V}}
\end{equation}
and
\begin{equation}
\label{eq28}
\ket{\chi^\gamma_{\text{E}}(t,\tau)}_{\text{V}} =  -\sqrt{\Gamma\Gamma_\perp}  \,\Theta(t-\tau)\int_{t_0}^{\tau}dt^\prime\,\xi(t^\prime) \,e^{-i H^g_{\text{V}} (t-\tau)}\exp\Big({i\Big( H^e_{\text{V}} + \frac{\Gamma+\Gamma_\perp}{2}\Big)(t^\prime-\tau)}\Big)e^{-i H^g_{\text{V}} t^\prime}\ket{\gamma}_{\text{V}}.
\end{equation}
Eq.~\eqref{eq26},\eqref{eq27},\eqref{eq28} are the solutions of the Schrodinger equation.

\subsection{Simplifying the vibration unitary}

The free vibration Hamiltonian $H^g_{\text{V}}$ and the T-V interaction Hamiltonian do not commute.
Therefore the unitary $\exp\Big({i\, H^e_{\text{V}}\,t }\Big)$ can be decomposed into a displacement operator $\mathcal{D}(\zeta)$, a rotation operator $\mathcal{R}(\theta)$ and some phase factors from the BCH expansion.
We shall use the following identity which is proved in \cite{Van_Brunt_2015} to decompose the vibration unitary,
\begin{equation}
    e^{x\,c^\dagger c+y\, c+y\, c^\dagger} =  e^{\frac{y}{x}((1-e^{-x})c-(1-e^{x})c^\dagger)} \,\,  e^{x \, c^\dagger c}\,\, e^{\frac{y^2}{x^2}\left(\frac{e^x-e^{-x}}{2}-x\right)},
\end{equation}
Using this identity, we can decompose the following operator,
\begin{equation}
\label{eq30}
    \exp\Big({i\, H^e_{\text{V}}\,t}\Big) = \underset{k}{\prod}\,\,\mathcal{D}_k(\zeta_k(t))\,\,\mathcal{R}_k(\Omega_k t) \,\,\exp\Big( i\phi_k(t)\Big),
\end{equation}
where, 
\begin{equation}
\zeta_k(t) = \frac{g_k}{\Omega_k}(1-e^{-i\Omega_k t}),
\end{equation}
\begin{equation}
    \phi_k(t) = \frac{\vert g_k\vert^2}{\Omega_k^2}\sin{(\Omega_k t)},
\end{equation}
and, $\mathcal{D}_k(x) = e^{x c_k- x^* {c_k}^\dagger}$, $\mathcal{R}_k(x) = e^{i x ({c_k}^\dagger c_k)}$.
Using this decomposition we can simplify the solutions to the Schrodinger equation further, obtaining
\begin{equation}
\label{eq35}
\begin{aligned}
    \ket{A_\gamma(t)}_{\text{V}} = \sqrt{\Gamma}\int_{t_0}^{t}dt^\prime\,\xi(t^\prime) \,\exp\Big({ \frac{\Gamma+\Gamma_\perp}{2}(t^\prime-t)}\Big)\underset{k}{\prod}\,e^{ i\phi_k(t^\prime-t)}\,\mathcal{D}_k(\zeta_k(t^\prime-t))\,\,\mathcal{R}_k(\Omega_k t) \,\ket{\gamma}_{\text{V}} \, ,
\end{aligned}
\end{equation}
\begin{equation}
\label{eq36}
\begin{aligned}
\ket{\chi^\gamma_{\text{P}}(t,\tau)}_{\text{V}} = \xi(\tau)e^{-i H^g_{\text{V}} t}\ket{\gamma}_{\text{V}}
- \Gamma & \,\Theta(t-\tau)\int_{t_0}^{\tau}dt^\prime\,\xi(t^\prime) \,\exp\Big({ \frac{\Gamma+\Gamma_\perp}{2}(t^\prime-\tau)}\Big)\\&
\underset{k}{\prod}\,e^{ i\phi_k(t^\prime-\tau)}\,\mathcal{R}_k(\Omega_k (\tau-t)) \,\mathcal{D}_k(\zeta_k(t^\prime-\tau))\,\,\mathcal{R}_k(-\Omega_k \tau) \,\ket{\gamma}_{\text{V}}
\end{aligned}
\end{equation}
and
\begin{equation}
\label{eq37}
\begin{aligned}
    \ket{\chi^\gamma_{\text{E}}(t,\tau)}_{\text{V}} =  -\sqrt{\Gamma\Gamma_\perp}  \,\Theta(t-\tau)\int_{t_0}^{\tau}dt^\prime\,&\xi(t^\prime) \,\exp\Big({ \frac{\Gamma+\Gamma_\perp}{2}(t^\prime-\tau)}\Big)\\&
\underset{k}{\prod}\,e^{ i\phi_k(t^\prime-\tau)}\,\mathcal{R}_k(\Omega_k (\tau-t)) \,\mathcal{D}_k(\zeta_k(t^\prime-\tau))\,\,\mathcal{R}_k(-\Omega_k \tau) \,\ket{\gamma}_{\text{V}} \, .
\end{aligned}
\end{equation}

\subsection{The state of the scattered pulse}
Now, we trace out the emitter, vibration and the electromagnetic degrees of freedom from the ansatz to obtain the state of the scattered pulse.
Note that the solution we have derived is for an arbitrary pure initial vibrational state $\ket{\gamma}_{\text{V}}$.
For a thermal initial state of the vibration, which can 
be written as $\sigma_\text{V}^{\text{th}}=\sum_\gamma \sigma_\gamma \ket{\gamma}_{\text{V}}\bra{\gamma}$, we use the linearity of the partial trace to obtain the following expressions,

\begin{equation}
\begin{aligned}
    \rho_{\text{P}}(t) = \text{Tr}_{\text{TVE}}\Big(\sum_\gamma \sigma_\gamma\ket{\psi_\gamma(t)}_{\text{TVPE}}\bra{\psi_\gamma(t)}\Big) = \Big(p_e(t)+p_{\Gamma_\perp}(t)\Big) \ket{0}_{\text{P}}\bra{0} + \rho^1_{\text{P}}(t),
\end{aligned}
\end{equation}

where,
\begin{equation}
\begin{aligned}
     p_e(t) = \sum_\gamma \sigma_\gamma\langle &A_\gamma(t)\vert A_\gamma(t)\rangle_{\text{V}}\\
    =\Gamma\int_{t_0}^{t}\int_{t_0}^{t}&dt^\prime dt^{\prime\prime}\,\xi(t^\prime)(\xi(t^{\prime\prime}))^* \,\exp\Big({ \frac{\Gamma+\Gamma_\perp}{2}(t^\prime++t^{\prime\prime}-2t)}\Big)\underset{k}{\prod}\,e^{ i\phi_k(t^\prime-t)-i\phi_k(t^{\prime\prime}-t)}\,\\&
\text{Tr}_{\text{V}}\Bigg(\mathcal{D}_k(\zeta_k(t^\prime-t))\,\,\mathcal{R}_k(\Omega_k t) \,\sum_\gamma \sigma_\gamma\ket{\gamma}_{\text{V}}\bra{\gamma}\mathcal{R}_k(-\Omega_k t)\mathcal{D}_k(-\zeta_k(t^{\prime\prime}-t))\Bigg)\\&
=\Gamma\int_{t_0}^{t}\int_{t_0}^{t}dt^\prime dt^{\prime\prime}\,\xi(t^\prime)(\xi(t^{\prime\prime}))^* \,\exp\Big({ \frac{\Gamma+\Gamma_\perp}{2}(t^\prime++t^{\prime\prime}-2t)}\Big)\\&\underset{k}{\prod}\,e^{ i\phi_k(t^\prime-t)-i\phi_k(t^{\prime\prime}-t)-i\Im(\zeta_k(t^\prime-t)\zeta_k(t^{\prime\prime}-t)^*)}\,
\text{Tr}_{\text{V}}\Bigg(\mathcal{D}_k(\zeta_k(t^\prime-t)-\zeta_k(t^{\prime\prime}-t))\,\sigma_\text{V}^{\text{th}}\Bigg),
\end{aligned}
\end{equation}
where $\Im(.)$ denotes the imaginary part of a complex number.
After simplifications,
\begin{equation}
\begin{aligned}
p_e(t) =  \Gamma\int_{t_0}^{t}\int_{t_0}^{t}dt^\prime dt^{\prime\prime}&\,\xi(t^\prime)(\xi(t^{\prime\prime}))^* \,\exp\Big({ \frac{\Gamma+\Gamma_\perp}{2}(t^\prime+t^{\prime\prime}-2t)}\Big) \\& \exp\Bigg(\sum_k \frac{\vert g_k\vert^2}{\Omega_k^2}\Bigg(\coth(\frac{\beta\Omega_k}{2})(\cos(\Omega_k(t^\prime-t^{\prime\prime})-1)+i\,\,\sin(\Omega_k(t^\prime-t^{\prime\prime})\Bigg)\Bigg) \,.
\end{aligned}
\end{equation}
Note, we have used the characteristics function of a thermal state which is defined as: $\text{Tr}(\mathcal{D}_k(x)\sigma_\text{V}^{\text{th}})= \exp(-\coth(\beta\Omega_k/2)\vert x\vert^2/2)$ where $\beta$ is the inverse temperature.
The probability of emission into the electromagnetic environment can be simplified to,
\begin{equation}
\begin{aligned}
    p_{\Gamma_\perp}(t) = \Gamma_\perp \int_{t_0}^{t} d\tau\, p_{e}(\tau).
\end{aligned}
\end{equation}

The scattered pulse can be written as
\begin{equation}
\begin{aligned}
    &\rho_{\text{P}}^1(t) =  \int\int d\tau d\tau^\prime\,\text{Tr}_{\text{TVE}}\Bigg(\sum_\gamma \sigma_\gamma \ket{\chi^\gamma_{\text{P}}(t,\tau)}_{\text{V}}\bra{\chi^\gamma_{\text{P}}(t,\tau^\prime)}\Bigg)\,a_{\text{P}}^\dagger(\tau)\ket{0}_{\text{P}}\bra{0}a_{\text{P}}(\tau^\prime)\\&
    = \int\int d\tau d\tau^\prime\,\,a_{\text{P}}^\dagger(\tau)\ket{0}_{\text{P}}\bra{0}a_{\text{P}}(\tau^\prime)\,\,\Bigg[\xi(\tau)(\xi(\tau^\prime))^* - \Gamma\xi(\tau^\prime)^*\Theta(t-\tau)\int_{t_0}^{\tau}dt^\prime\,\xi(t^\prime) \,\exp\Big({ \frac{\Gamma+\Gamma_\perp}{2}(t^\prime-\tau)}\Big)\\&
\underset{k}{\prod}\,e^{ i\phi_k(t^\prime-\tau)}\,\text{Tr}_{\text{V}}\Bigg(\mathcal{R}_k(\Omega_k (\tau-t)) \,\mathcal{D}_k(\zeta_k(t^\prime-\tau))\,\,\mathcal{R}_k(-\Omega_k (\tau-t)) \,\sum_\gamma \sigma_\gamma\ket{\gamma}_{\text{V}}\bra{\gamma}\Bigg)\\&-\Gamma\xi(\tau)\Theta(t-\tau^\prime)\int_{t_0}^{\tau^\prime}dt^\prime\,\xi(t^\prime)^* \,\exp\Big({ \frac{\Gamma+\Gamma_\perp}{2}(t^\prime-\tau^\prime)}\Big)\\&
\underset{k}{\prod}\,e^{ -i\phi_k(t^\prime-\tau^\prime)}\,\text{Tr}_{\text{V}}\Bigg(\mathcal{R}_k(-\Omega_k (\tau^\prime-t)) \,\mathcal{D}_k(-\zeta_k(t^\prime-\tau^\prime))\,\,\mathcal{R}_k(\Omega_k (\tau^\prime-t)) \,\sum_\gamma \sigma_\gamma\ket{\gamma}_{\text{V}}\bra{\gamma}\Bigg)\\&
+\Gamma^2 \Theta(t-\tau)\Theta(t-\tau^\prime)\int_{t_0}^{\tau}\int_{t_0}^{\tau^\prime}dt^\prime dt^{\prime\prime}\,\xi(t^\prime)\xi(t^{\prime\prime})^* \,\exp\Big({ \frac{\Gamma+\Gamma_\perp}{2}(t^\prime+t^{\prime\prime}-\tau-\tau^\prime)}\Big)
\underset{k}{\prod}\,e^{ i\phi_k(t^{\prime}-\tau)-i\phi_k(t^{\prime\prime}-\tau^\prime)}\,\\&
\text{Tr}_{\text{V}}\Bigg(\mathcal{R}_k(\Omega_k (\tau-t)) \,\mathcal{D}_k(\zeta_k(t^\prime-\tau))\,\mathcal{R}_k(-\Omega_k (\tau)) \,\sum_\gamma \sigma_\gamma\ket{\gamma}_{\text{V}}\bra{\gamma}\mathcal{R}_k(\Omega_k (\tau^\prime))\mathcal{D}_k(-\zeta_k(t^{\prime\prime}-\tau^\prime))\mathcal{R}_k(-\Omega_k (\tau^\prime-t))\Bigg)\Bigg].
\end{aligned}
\end{equation}
After simplifications, we obtain
\begin{equation}
\begin{aligned}
    \rho_{\text{P}}^1(t)& 
    = \int\int d\tau d\tau^\prime\,\,a_{\text{P}}^\dagger(\tau)\ket{0}_{\text{P}}\bra{0}a_{\text{P}}(\tau^\prime)\,\,\Bigg[\xi(\tau)(\xi(\tau^\prime))^* \\& - \Gamma\xi(\tau^\prime)^*\Theta(t-\tau)\int_{t_0}^{\tau}dt^\prime\,\xi(t^\prime) \,\exp\Big({ \frac{\Gamma+\Gamma_\perp}{2}(t^\prime-\tau)}\Big)
\underset{k}{\prod}\,e^{ i\phi_k(t^\prime-\tau)}\,\text{Tr}_{\text{V}}\Bigg(\mathcal{D}_k(\zeta_k(t^\prime-\tau))\,\sigma_\text{V}^{\text{th}}\Bigg)\\&-\Gamma\xi(\tau)\Theta(t-\tau^\prime)\int_{t_0}^{\tau^\prime}dt^\prime\,\xi(t^\prime)^* \,\exp\Big({ \frac{\Gamma+\Gamma_\perp}{2}(t^\prime-\tau^\prime)}\Big)
\underset{k}{\prod}\,e^{ -i\phi_k(t^\prime-\tau^\prime)}\,\text{Tr}_{\text{V}}\Bigg(\mathcal{D}_k(-\zeta_k(t^\prime-\tau^\prime))\,\sigma_\text{V}^{\text{th}}\Bigg)\\&
+\Gamma^2 \Theta(t-\tau)\Theta(t-\tau^\prime)\int_{t_0}^{\tau}\int_{t_0}^{\tau^\prime}dt^\prime dt^{\prime\prime}\,\xi(t^\prime)\xi(t^{\prime\prime})^* \,\exp\Big({ \frac{\Gamma+\Gamma_\perp}{2}(t^\prime+t^{\prime\prime}-\tau-\tau^\prime)}\Big)
\underset{k}{\prod}\,e^{ i\phi_k(t^{\prime}-\tau)-i\phi_k(t^{\prime\prime}-\tau^\prime)}\,\\&
\text{Tr}_{\text{V}}\Bigg(\mathcal{D}_k(-\zeta_k(t^{\prime\prime}-\tau^\prime))\,\mathcal{D}_k(\zeta_k(t^\prime-\tau)e^{i\Omega_k(\tau-\tau^\prime)}) \,\sigma_\text{V}^{\text{th}}\Bigg)\Bigg].
\end{aligned}
\end{equation}
Simplifying further using the characteristic function, we obtain
\begin{equation}
\label{eq44}
\begin{aligned}
    \rho_{\text{P}}^1(t)& 
    = \int\int d\tau d\tau^\prime\,\,a_{\text{P}}^\dagger(\tau)\ket{0}_{\text{P}}\bra{0}a_{\text{P}}(\tau^\prime)\,\,\Bigg[\xi(\tau)(\xi(\tau^\prime))^*  - \Gamma\xi(\tau^\prime)^*\Theta(t-\tau)\int_{t_0}^{\tau}dt^\prime\,\xi(t^\prime) \,\exp\Big({ \frac{\Gamma+\Gamma_\perp}{2}(t^\prime-\tau)}+
 \Lambda_1(t^\prime-\tau)\Big)\\&-\Gamma\xi(\tau)\Theta(t-\tau^\prime)\int_{t_0}^{\tau^\prime}dt^\prime\,\xi(t^\prime)^* \,\exp\Big({ \frac{\Gamma+\Gamma_\perp}{2}(t^\prime-\tau^\prime)}+
\Lambda_1(-t^\prime+\tau^\prime)\Big)\\&
+\Gamma^2 \Theta(t-\tau)\Theta(t-\tau^\prime)\int_{t_0}^{\tau}\int_{t_0}^{\tau^\prime}dt^\prime dt^{\prime\prime}\,\xi(t^\prime)\xi(t^{\prime\prime})^* \,\exp\Big({ \frac{\Gamma+\Gamma_\perp}{2}(t^\prime+t^{\prime\prime}-\tau-\tau^\prime)}
\,+\Lambda_2(t^\prime,t^{\prime\prime},\tau,\tau^\prime)\Big)\Bigg].
\end{aligned}
\end{equation}
where
\begin{equation}
\label{eq45}
\Lambda_1(t) = \sum_k \frac{\vert g_k\vert^2}{\Omega_k^2}\Bigg(\coth(\frac{\beta\Omega_k}{2})(\cos(\Omega_k t)-1)+i\,\,\sin(\Omega_k t)\Bigg),
\end{equation}
and
\begin{equation}
    \Lambda_2(t_1,t_2,\tau,\tau^\prime) = \Lambda_1(t_1-\tau)+\Lambda_1(t_2-\tau^\prime)^*-\Lambda_1(t_1-\tau^\prime) - \Lambda_1(t_2-\tau)^* + \Lambda_1(t_1-t_2) + \Lambda_1(\tau-\tau^\prime).
\end{equation}

The TLS-vibration coupling can be described by a continuous spectral density $J(\Omega)$, which for the discrete case takes the following form:
\begin{equation}
    J(\Omega) = \sum_k \vert g_k\vert^2 \delta(\Omega-\Omega_k) .
\end{equation}
Using this spectral density we can rewrite the vibration propagator as
\begin{equation}
\label{eq48}
\Lambda_1(t) = \int_0^\infty d\Omega\,\, \frac{J(\Omega)}{\Omega^2}\Bigg(\coth(\frac{\beta\Omega}{2})(\cos(\Omega t)-1)+i\,\,\sin(\Omega t)\Bigg).
\end{equation}

At $t\gg(\Gamma+\Gamma_\perp)^{-1}$, Eq.~\eqref{eq44} simplifies to Eq.~\eqref{scatterp} in the main text.

\section{Numerical evaluation of the temporal density matrices (TDM)}
\label{asec3}
In this section, we discuss how to numerically evaluate the TDM in Eq.~\eqref{scatterp} in a $N\times N$ grid of time-bins: $(\tau,\tau^\prime)\in [\tau_0, \tau_{N-1}]\times[\tau_0, \tau_{N-1}]$. The upper and lower limits of the time-bins, i.e., $\tau_0$ and $\tau_{N-1}$ are chosen according to the largest time scale of the dynamics, i.e., the emitter lifetime $1/(\Gamma+\Gamma_\perp)$. The bin-size $\Delta_\tau\equiv(\tau_{N-1}-\tau_0)/N$ are chosen respecting the smallest time-scale of the dynamics which in this case, is the frequency of the vibration: $\Omega_0^{-1}$. The single convolutions of Eq.~\eqref{eq44} can be straightforwardly evaluated by using the convolution theorem and fast F.T. of the discretized incident pulse $\xi(\tau)$ and the emitter response function $e^{ \frac{\Gamma+\Gamma_\perp}{2}\tau+
 \Lambda_1(\tau)}$. The double integral in the TDM is not expressed in a double convolution form due to the non-stationary terms of $\Lambda_2$ in Eq.~\eqref{lm2}. However, for a fixed value of $\tau-\tau^\prime$ the integrals become a double convolution. This corresponds to a specific set of off-diagonal elements. Thus we run double convolutions to evaluate each set of off-diagonal elements of the density matrix. This step incurs a computational cost scaling of $\mathcal{O}(N^3 \text{log}(N))$. 
 
 In contrast to the main text, we choose a different vibrational mode with lower frequency $\Omega_0=100\text{cm}^{-1}$ to see more exaggerated thermal effects (see Fig.~\ref{fig:qfi100}). This mode may physically corresponds to the delocalised vibrations of molecules \cite{BoehmkeAmoruso2024}. This choice along with $0\leq\lambda_0\leq 1$ allows us to faithfully capture all timescales of our system with $N=1000$. The full list of parameter values used for the supplementary note is given in Table~\ref{tab:1}.

The derivatives of the TDM with respect to the emitter parameters do not alter the convolution structure of the integrals, thus can be numerically evaluated in the same way.
The parameter values used for this computation are given in the Table.~\ref{tab:1}.

\begin{table}
    \centering
    \begin{tabular}{c@{\hspace{1cm}}c@{\hspace{1cm}}c@{\hspace{1cm}}c@{\hspace{1cm}}c@{\hspace{1cm}}c@{\hspace{1cm}}c}
\hline\hline
$\Gamma$ &
$\Omega_0$ &
$\tau_0$ &
$\tau_{N-1}$ &
$N$ &
$\gamma$ &
$T_\sigma$ \\
\hline
$0.15\,\mathrm{THz}$ &
$100\,\mathrm{cm}^{-1}$ &
$-10/\Gamma$ &
$10/\Gamma$ &
$1000$ &
$10\,\mathrm{cm}^{-1}$ &
$1/\Gamma$ \\
\hline\hline
\end{tabular}
    \caption{Parameter values used for the numerical evaluation of the TDM and QFI. The cutoff frequency $\gamma$ is applied to the continuum vibrational spectral densities discussed in Sec.~\ref{asec9}.}
    \label{tab:1}
\end{table}

\section{Jacobi-Anger expansion and the SDM of the scattered pulse}
\label{asec4}
We use the following identity which can be derived from the Jacobi-Anger expansion to transform the time-domain wavefunction of the scattered pulse,
\begin{eqnarray}
\label{eq1}
    \exp(a \cos \theta + i b \sin \theta) = \sum_{k=-\infty}^{\infty} a_k e^{i k \theta},
\end{eqnarray}
where the coefficients $a_k$ are given by:
\begin{eqnarray}
    a_k = \sum_{n=-\infty}^{\infty} I_n(a) J_{k-n}(b) = I_n(\sqrt{a^2-b^2})\left(\frac{a-b}{a+b}\right)^{n/2}.
\end{eqnarray}
Here $J_n(z)$ and $I_n(z)$ are the Bessel functions and modified Bessel functions of first kind of order n and variable $z$. The dephasing function for a single mode vibration takes the following form,  
\begin{eqnarray}
    \Lambda_1(t) = \lambda_0\bar{n}(\cos(\Omega_0 t)-1)
    +i\,\sin(\Omega_0 t).
\end{eqnarray}
Thus we can write the following using the identity in Eq.~\eqref{eq1},
\begin{eqnarray}
\label{eq4}
    \exp(\Lambda_1(t)) = \sum_{k=-\infty}^{\infty} f_k e^{i k \Omega_0 t}
\end{eqnarray}
where,
\begin{eqnarray}
    f_k = e^{-\lambda_0  \bar{n}} I_k\left(\lambda_0 \sqrt{\bar{n}^2-1}\right)\left(\frac{\bar{n}+1}{\bar{n}-1}\right)^{k/2}.
\end{eqnarray}
Correspondingly we can write the following,
\begin{eqnarray}
\label{eq6}
    \exp(-\Lambda_1(t)) = \sum_{k=-\infty}^{\infty} d_k e^{i k \Omega_0 t}
\end{eqnarray}
where,
\begin{eqnarray}
    d_k = e^{\lambda_0  \bar{n}} I_k\left(-\lambda_0 \sqrt{\bar{n}^2-1}\right)\left(\frac{\bar{n}+1}{\bar{n}-1}\right)^{k/2}.
\end{eqnarray}

Using Eq.~\eqref{eq4}, Eq.~\eqref{eq6} we can derive the following,

\begin{eqnarray}
\label{eq11}
    e^{\Lambda_2(t_1,t_2,\tau,\tau^\prime)} = \sum_{k_1,...,k_6=-\infty}^{\infty}f_{k_1}f_{k_2}d_{k_3}d_{k_4}f_{k_5}f_{k_6}\exp\Big(i\Omega_0 \big( (t_1-\tau)(k_1-k_4-k_6)
    \nonumber\\-(t_2-\tau^\prime)(k_2-k_3-k_6) + (t_1-t_2)(k_3+k_4+k_5+k_6)
    \big)\Big)\\
    =  \sum_{k,l,m=-\infty}^{\infty} C^l_{mn} \exp\Big(i\Omega_0 \big( m(t_1-\tau)-n(t_2-\tau^\prime) + l(t_1-t_2)
    \big)\Big)\nonumber
\end{eqnarray}
where, $l= k_3+k_4+k_5+k_6$, $m=k_1-k_4-k_6$, $n=k_2-k_3-k_6$, and,
\begin{eqnarray}
    C^l_{mn} = \sum_{k_3,k_4,k_6=-\infty}^{\infty} f_{m+k_4+k_6}f_{n+k_3+k_6}d_{k_3}d_{k_4}f_{l-k_3-k_4-k_6}f_{k_6}.
\end{eqnarray}

\subsection{Scattered pulse spectral density matrix (SDM)}

Now we transform the time domain wave-packet of the scattered pulse to the frequency domain. We define the frequency domain wave-packet $\tilde{\varrho}(\omega_1,\omega_2)$ as the 2d Hermitian F.T. of the time-domain wave-packet $\varrho(\tau,\tau^\prime)$ where the 2d Hermitian F.T. means F.T. along the rows and inverse-F.T. along the columns which preserves the Hermiticity of the density matrix. Correspondingly the inverse Hermitian F.T. is defined as the inverse F.T. along the rows and F.T. along the columns. We also set the initial time $t_0\rightarrow\infty$ and the start of the experiment is set by the arrival of the pulse $\xi(t)$. Thus, we rewrite the time domain wave-packet as,
\begin{equation}
\label{s50}
\begin{aligned}
    \varrho_{\text{P}}(\tau,\tau^\prime)
= \varrho_0(\tau,\tau^\prime) - \varrho_1(\tau,\tau^\prime) - \varrho_1(\tau^\prime,\tau)^* + \varrho_2(\tau,\tau^\prime)
\end{aligned}
\end{equation}
where we decomposed the contributions of the incident pulse, the single and double integrals respectively with the functions $\varrho_0,\varrho_1$ and $\varrho_2$. $\varrho_1$ writes a convolution integral,
\begin{eqnarray}
\label{eq13}
    \tilde{\varrho_1}(\omega_1,\omega_2) = \Gamma \sqrt{2\pi}\,\tilde{\xi}(\omega_1)\tilde{\xi}(\omega_2)^* \mathcal{F}_t\left[\Theta(t)\exp\left(-\frac{(\Gamma+\Gamma_\perp) t}{2}+\Lambda_1(-t)\right)\right](\omega_1)
\end{eqnarray}
where $\mathcal{F}_t\left[\cdot\right]$ denotes the F.T. Thus, in the frequency domain using Eq.~\eqref{eq4},
\begin{eqnarray}
    \tilde{\varrho_1}(\omega_1,\omega_2) = 4\Gamma\sum_{k=-\infty}^{\infty} \frac{ f_k }{\Gamma+\Gamma_\perp-2i(\omega_1-k\Omega)}\,\tilde{\xi}(\omega_1)\tilde{\xi}(\omega_2)^*
\end{eqnarray}

Similarly, using Eq.~\eqref{eq11} we transform $\varrho_2(\tau,\tau^\prime)$ in the frequency domain which obtains the following,
\begin{equation}
\begin{aligned}
    \tilde{\varrho_2}(\omega_1,\omega_2) = 2\pi\Gamma^2  \sum_{l,m,n=-\infty}^{\infty} C^l_{mn} \mathcal{F}_{t_1}\left[\xi(t_1)e^{i\Omega l t_1}\right](\omega_1)\,\,\mathcal{F}^{-1}_{t_2}\left[\xi(t_2)^*e^{-i\Omega l t_2}\right](\omega_2)\\
    \mathcal{F}_{\tau}\left[\Theta(\tau)\exp\left(-\frac{(\Gamma +\Gamma_\perp)\tau}{2}+\Lambda_1(-\tau)\right)\right](\omega_1)\,\, \mathcal{F}^{-1}_{\tau^\prime}\left[\Theta(\tau^\prime)\exp\left(-\frac{(\Gamma +\Gamma_\perp) \tau^\prime}{2}+\Lambda_1(-\tau^\prime)^*\right)\right](\omega_2)\\
    = 4\Gamma^2  \sum_{l,m,n=-\infty}^{\infty} C^l_{mn} \,\,\frac{\tilde{\xi}(\omega_1+ l\Omega)\tilde{\xi}(\omega_2+ l\Omega)^*}{\left(\Gamma+\Gamma_\perp-2i(\omega_1-m\Omega)\right)\left(\Gamma+\Gamma_\perp+2i(\omega_1-n\Omega)\right)}
\end{aligned}
\end{equation}
Putting these together in Eq.~\eqref{s50}, in the frequency domain we express the spectral density matrix of the scattered pulse as follows,
\begin{eqnarray}
\label{s54}
    \tilde{\varrho}_\text{P}(\omega_1,\omega_2) = \tilde{\xi}(\omega_1)\tilde{\xi}(\omega_2)^* - 4\Gamma \, \sum_{k=-\infty}^{\infty} \left[\frac{ f_k \tilde{\xi}(\omega_1)\tilde{\xi}(\omega_2)^*\left(\Gamma+\Gamma_\perp-i(\omega_1-\omega_2)\right)}{\left(\Gamma+\Gamma_\perp-2i(\omega_1-k\Omega)\right)\left(\Gamma+\Gamma_\perp+2i(\omega_2-k\Omega)\right)}\right]\\ \nonumber
    +4\Gamma^2  \sum_{l,m,n=-\infty}^{\infty} C^l_{mn} \,\,\frac{\tilde{\xi}(\omega_1+ l\Omega)\tilde{\xi}(\omega_2+ l\Omega)^*}{\left(\Gamma+\Gamma_\perp-2i(\omega_1-m\Omega)\right)\left(\Gamma+\Gamma_\perp+2i(\omega_2-n\Omega)\right)}.
\end{eqnarray}

\section{The spectra of the scattered pulse}
\label{asec5}
The spectra of the scattered pulse which are defined as the diagonal elements of the SDM observed in the FRP can be derived from Eq.~\eqref{s54} which take the following form in the limit $\Gamma\ll\Omega_0$ and $T_\sigma\Omega_0\gg 
 1$,
\begin{equation}
\label{eq9}
\begin{split}
    \mathcal{S}(\omega) \equiv \tilde{\varrho}_\text{P}(\omega,\omega) = \vert \tilde{\xi}(\omega)\vert^2 
    - \sum_{k}\frac{4 f_k \Gamma(\Gamma+\Gamma_\perp)\vert \tilde{\xi}(\omega)\vert^2 }{(\Gamma+\Gamma_\perp)^2+4(\omega-k\Omega_0)^2} 
    + \sum_{k}\frac{4\Gamma^2 C^k_{k,k}\vert\tilde{\xi}(\omega- k\Omega_0)\vert^2}{(\Gamma+\Gamma_\perp)^2+4(\omega-k\Omega_0)^2}.
\end{split}
\end{equation}
The first element in this expression denotes the probability of the pulse propagating through the emitter without interaction. The coefficients $f_k$ and $C^k_{k,k}$ are positive. This ensures the second term in Eq.~\eqref{eq9} is a negative contribution to the spectrum, attributable to the absorption events ($\mathcal{S}_{\text{absorption}}(\omega)$), while the third term is a positive contribution, stemming from the emission events ($\mathcal{S}_{\text{emission}}(\omega)$). Thus, the spectrum $\mathcal{S}(\omega)$ could be written in a more physically meaningful form,
\begin{equation}
    \mathcal{S}(\omega) = \mathcal{S}_{\text{input}}(\omega) - \mathcal{S}_{\text{absorption}}(\omega) + \mathcal{S}_{\text{emission}}(\omega).
\end{equation}

\section{Derivation of $\mathcal{Q}_\text{bound}$}
\label{asec6}
Due to the data-processing inequality, the QFI of the pulse-state $\rho_\text{P}(\infty)$ is always upper-bounded by the QFI of the pulse-vibration joint state $\rho_\text{PV}(\infty)$:
\begin{eqnarray}
    \mathcal{Q}\left(\rho_\text{P}(\infty)\right) \leq\mathcal{Q}\left(\rho_\text{PV}(\infty)\right).
\end{eqnarray}
With $\Gamma_\perp=0$, $t\gg \Gamma^{-1}$, and, a pure vibrational initial state $\ket{\gamma}_\text{V}$, the pulse-vibration joint state is a pure state: $\rho_\text{PV}(\infty)=\ket{1_\gamma(\infty)}\bra{1_\gamma(\infty)}_{\text{PV}}$ which is given by the wave-packet in Eq.~\eqref{eq36}. Therefore, the QFI of this state for estimating $\Gamma$ can be written as:
\begin{eqnarray}
    \mathcal{Q}\left(\ket{1_\gamma(\infty)}_{\text{PV}}\right) = 4 \Big[\vert\vert\partial_\Gamma\ket{1_\gamma(\infty)}_{\text{PV}}\vert\vert^2 -\vert \bra{1_\gamma(\infty)}\partial_\Gamma\ket{1_\gamma(\infty)}_{\text{PV}}\vert^2\Big],
\end{eqnarray}
where $\vert\vert\ket{\psi}\vert\vert^2= \langle \psi\ket{\psi}$. The second term is a negative contribution, thus, the first term serves as an upper-bound to the QFI. In addition to that, the QFI satisfies the following extended convexity:
\begin{eqnarray}
    \mathcal{Q}\Big(\sum_k q_k \rho_k \Big) \leq \sum_k  q_k\mathcal{Q}\big( \rho_k \big),
\end{eqnarray}
assuming $q_k$ are not dependent on the parameter. Therefore, for a thermal initial state, we can straightforwardly set the following upper-bound:
\begin{eqnarray}
    \mathcal{Q}\left(\rho_\text{P}(\infty)\right) \leq\mathcal{Q}\left(\rho_\text{PV}(\infty)\right) = \mathcal{Q}\left(\sum_\gamma \sigma_\gamma \ket{1_\gamma(\infty)}\bra{1_\gamma(\infty)}_{\text{PV}}\right) \leq 4 \sum_\gamma \sigma_\gamma \vert\vert\partial_\Gamma\ket{1_\gamma(\infty)}_{\text{PV}}\vert\vert^2 = \mathcal{Q}_\text{bound}.
\end{eqnarray}
Using Eq.~\eqref{eq36} one can derive the following:
\begin{equation}
    \mathcal{Q}_\text{bound} = 4\int_{-\infty}^{\infty}d\tau\int_{-\infty}^{\tau}\int_{-\infty}^{\tau}dt^\prime dt^{\prime\prime} \xi(t^\prime)\xi(t^{\prime\prime})^*\left(1-\frac{\Gamma}{2}(t^\prime-\tau)\right)\left(1-\frac{\Gamma}{2}(t^{\prime\prime}-\tau)\right)\exp\left({{ \frac{\Gamma}{2}(t^\prime+t^{\prime\prime}-2\tau)}
\,+\Lambda_1(t^\prime-t^{\prime\prime})}\right).
\end{equation}
For a single-mode vibration, this expression can be simplified in the frequency domain:
\begin{eqnarray}
    \mathcal{Q}_{\text{bound}} = \sum_{k=-\infty}^{\infty} f_k \int_{-\infty}^{\infty} d\omega \left(\frac{64(\omega-k\Omega_0)^2}{(\Gamma^2+4(\omega-k\Omega_0)^2)^2} \right)\vert\tilde{\xi}(\omega)\vert^2.
\end{eqnarray}

\begin{figure}
    \centering
    \includegraphics[width=0.8\linewidth]{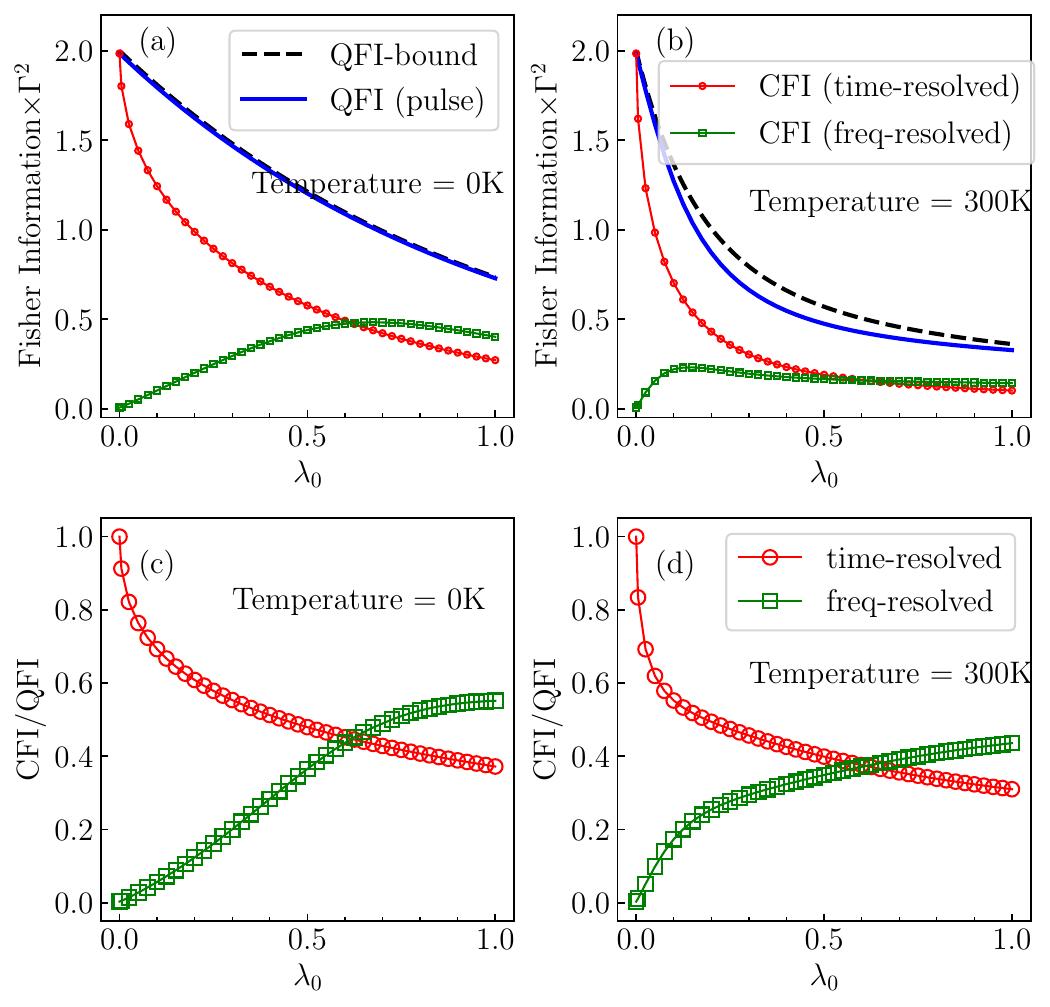}
    \caption{Fisher information (in the units of $\Gamma^2$) of estimating $\Gamma$ from the scattered pulse as a function of the Huang-Rhys factor $\lambda_0$.  The curves display the QFI upper-bound $\mathcal{Q}_{\text{bound}}$ (black dashed), QFI of the scattered pulse $\mathcal{Q}(\rho_{\text{P}}(\infty))$ (blue solid) and CFI for time- (red small circles) and frequency-resolved (green small squares) photon counting. (c,d) plots the corresponding ratios of the CFI and QFI of the scattered pulse. The incident pulse-shape is a decaying exponential (in the time-domain) $\xi(t)= \exp(-t/(2T_\sigma))\Theta(t)/\sqrt{T_\sigma}$ with pulse-duration $T_\sigma=1/\Gamma$. We evaluate this state numerically via the convolution theorem \cite[Sec.~\ref{asec3}]{supp}. The system parameters are: $\Gamma=0.15\text{THz}$, $\Omega_0=100\text{cm}^{-1}$.}
\label{fig:qfi100}
\end{figure}

Fig.~\ref{fig:qfi100} shows the Fisher information as a function of $\lambda_0$ for $\Omega_0 =100~\text{cm}^{-1}$. At room temperature, the average vibrational excitation number is $\bar{n} \approx 4.25$, so temperature-dependent effects are significant. In particular, the QFI decreases more rapidly with increasing $\lambda_0$ at room temperature than at zero temperature. Moreover, the QFI bound is less tight at room temperature because it relies on the extended convexity property of the QFI. 

At typical vibrational frequencies ($\gtrsim 1000~\text{cm}^{-1}$), this effect becomes dormant at room temperature, because the thermal population is similar to that for the zero temperature, as discussed in the main text. Consequently, higher temperatures are required for the effect to become observable.

\section{Fisher informations for estimating the Huang-Rhys factor}
\label{asec7}

\begin{figure}
    \centering
    \includegraphics[width=0.8\linewidth]{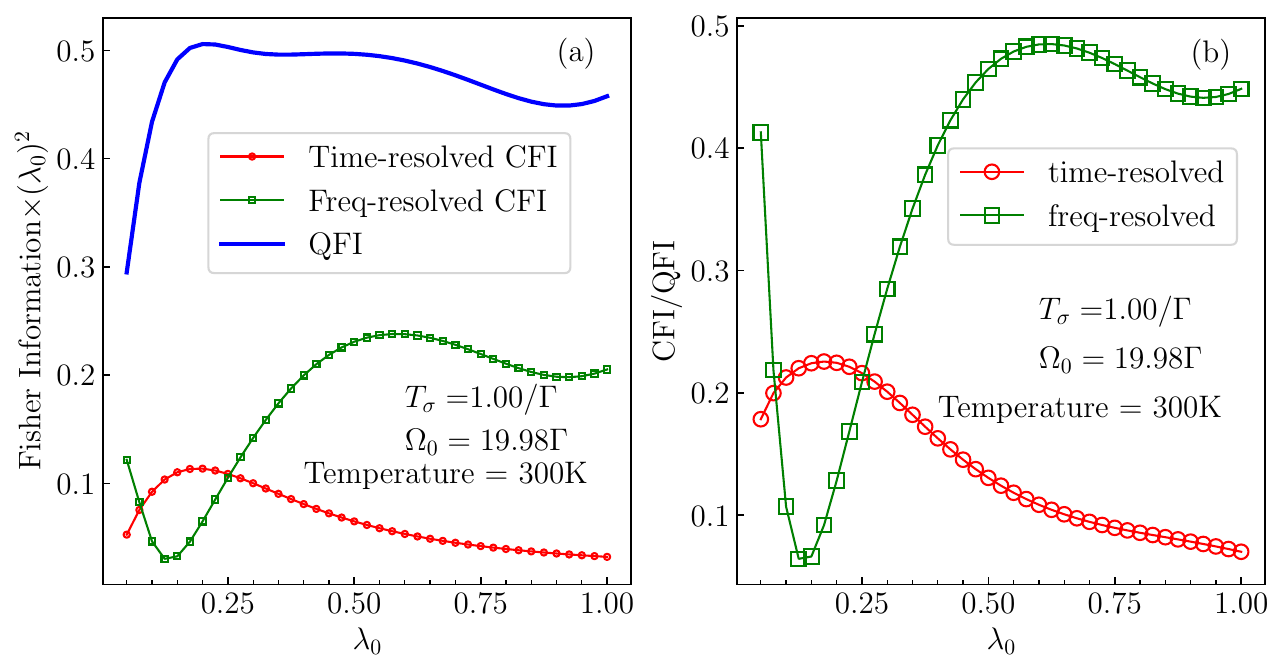}
    \caption{(a) Fisher information (in units of $(\lambda_0)^2$) contained in the scattered pulse $\rho_\text{P}(\infty)$ for estimating the Huang–Rhys factor $\lambda_0$. The curves show the QFI (blue solid) and the CFI for time-resolved (red small circles) and frequency-resolved (green small squares) photon-counting measurements. (b) Ratio of CFI to QFI for these measurement schemes. The incident pulse is an exponentially decaying pulse with duration $T_\sigma=1/\Gamma$, $\Gamma=0.15$THz, $\Omega_0=100\text{cm}^{-1}$, temperature = $300$K. }
    \label{figqfilmd0}
\end{figure}

Fig.~\ref{figqfilmd0}(a) shows the QFI of the scattered pulse for estimating the Huang--Rhys factor $\lambda_0$, together with the CFI obtained from time-resolved and frequency-resolved photon-counting measurements. For small values of $\lambda_0$ (up to $\sim 0.1$), the QFI increases. In the same regime, the frequency-resolved CFI decreases, whereas the time-resolved CFI increases (see also Fig.~\ref{figqfilmd0}(b)). This behavior is primarily governed by the suppression of the zero-phonon line (Rayleigh scattering) as the excited state population is redistributed into higher vibrational levels. At larger values of $\lambda_0$, this population redistribution becomes more pronounced, leading to the observed changes in both the QFI and CFI. Importantly, these simple photon-counting measurements can, in certain parameter regimes, extract nearly half of the QFI in the scattered pulse.

\section{Linewidth estimation with imperfect detector}
\label{asec_imperfect}
\begin{figure}
    \centering
    \includegraphics[width=\linewidth]{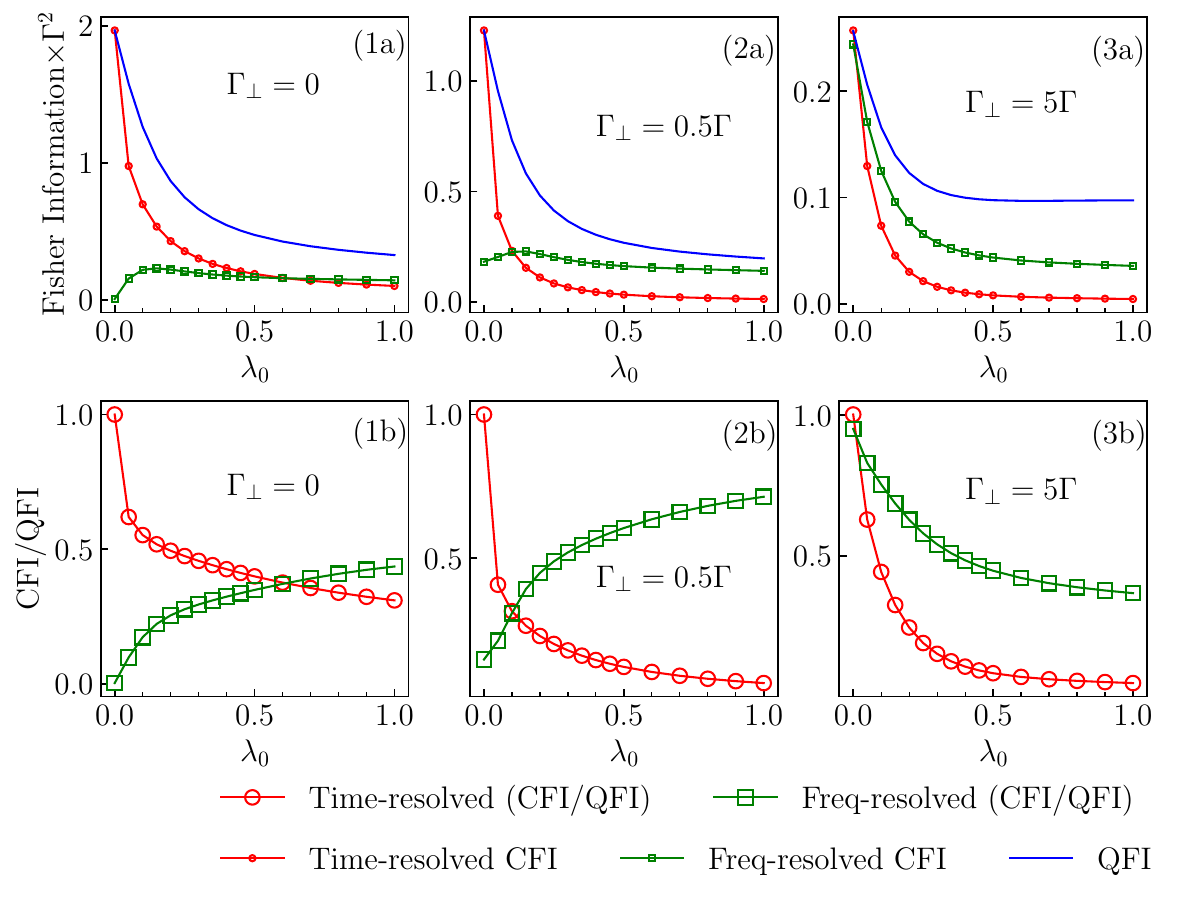}
    \caption{Fisher information (in units of $\Gamma^2$) contained in the scattered pulse $\rho_\text{P}(\infty)$ for estimating the linewidth $\Gamma$ with an imperfect detectors ($\Gamma_\perp>0$). The curves in the upper-panel (1a,2a,3a) show the QFI (blue solid) and the CFI for time-resolved (red small circles) and frequency-resolved (green small squares) photon-counting measurements. In the lower-panel (1b,2b,3b), the curves display the Ratio of CFI to QFI for these measurement schemes. The values for the detectors are chosen to be $\Gamma_\perp=0.0$ (1a,1b), $\Gamma_\perp=0.5\Gamma$ (2a,2b), and, $\Gamma_\perp=5\Gamma$ (3a,3b).  The incident pulse is an exponentially decaying pulse with duration $T_\sigma=1/\Gamma$, $\Gamma=0.15$THz, $\Omega_0=100\text{cm}^{-1}$, temperature = $300$K. }
    \label{fig:lossy}
\end{figure}
In this section, we compute the Fisher information for $\Gamma$ estimation when the photo-detectors are imperfect, i.e., $\Gamma_\perp\neq 0$. In this case, the asymptotic state of the scattered pulse is given by,
\begin{eqnarray}
    \rho_{\text{P}}(\infty) = p_{\Gamma_\perp} \ket{0}_{\text{P}}\bra{0} + \rho^1_{\text{P}}(\infty),
\end{eqnarray}
where $\rho^1_{\text{P}}(\infty)$ is the unnormalised state of the scattered single-photon for which $\text{Tr}(\rho^1_{\text{P}}(\infty))=1-p_{\Gamma_\perp}$. The SLD for $\Gamma$ estimation for this state takes the following form,
\begin{eqnarray}
    L_\Gamma = \frac{\partial_\Gamma p_{\Gamma_\perp}}{p_{\Gamma_\perp}}\ket{0}_{\text{P}}\bra{0} + L_\Gamma^1,
\end{eqnarray}
where $L_\Gamma^1$ is the solution of the Lyapunov equation for the unnormalised state: $L_\Gamma^1\rho^1_{\text{P}}(\infty)+ \rho^1_{\text{P}}(\infty)L_\Gamma^1 = 2\partial_\Gamma\rho^1_{\text{P}}(\infty)$. Thus, the total QFI can be written as:
\begin{eqnarray}
    \mathcal{Q}(\rho_{\text{P}}(\infty)) \equiv \text{Tr}\left(\rho_{\text{P}}(\infty)(L_\Gamma)^2\right) = \frac{\left(\partial_\Gamma p_{\Gamma_\perp}\right)^2}{p_{\Gamma_\perp}} + \text{Tr}\left(\rho^1_{\text{P}}(\infty)(L_\Gamma^1)^2\right).
\end{eqnarray}
The set of POVM for the TRP is now given by: $\{\Pi_\tau\}\equiv\{\ket{0}_{\text{P}}\bra{0}, \ket{1_\tau}_\text{P}\bra{1_\tau}\}$ and for FRP: $\{\Pi_\omega\}\equiv\{\ket{0}_{\text{P}}\bra{0}, \ket{1_\omega}_\text{P}\bra{1_\omega}\}$. Thus, the CFI for TRP is given by:
\begin{eqnarray}
    \mathcal{C}\left(\rho_{\text{P}}(\infty),\{\Pi_\tau\}\right) = \frac{\left(\partial_\Gamma p_{\Gamma_\perp}\right)^2}{p_{\Gamma_\perp}} + \int_{t_0}^{\infty} d\tau \frac{\left(\partial_\Gamma \varrho_\text{P}(\tau,\tau)\right)^2}{\varrho_\text{P}(\tau,\tau)},
\end{eqnarray}
and for FRP:
\begin{eqnarray}
\label{eqs67}
    \mathcal{C}\left(\rho_{\text{P}}(\infty),\{\Pi_\omega\}\right) = \frac{\left(\partial_\Gamma p_{\Gamma_\perp}\right)^2}{p_{\Gamma_\perp}} + \int_{-\infty}^{\infty} d\omega \frac{\left(\partial_\Gamma \mathcal{S}(\omega)\right)^2}{\mathcal{S}(\omega)}.
\end{eqnarray}
Using these relations, we compute the Fisher informations for various values of $\Gamma_\perp$ in Fig.~\ref{fig:lossy}. At $\lambda_0=0$, the TRP is optimal (Fig.~\ref{fig:lossy}(1b,2b,3b)) which is consistent with the previous results \cite{albarelli2023fundamental}. However, for $\Gamma_\perp>0$, the FRP-CFI is nonzero, due to the contributions coming from the photonic losses in Eq.~\eqref{eqs67}. The conclusion that TRP performs better than FRP for lower vibrational coupling strengths and vice-versa holds true for imperfect detectors. Interestingly, the cross-over of their sub-optimality happen at lower coupling strengths for higher $\Gamma_\perp$ (see Fig.~\ref{fig:lossy}(1b,2b,3b)).

\section{The vibration propagator for continuous spectral densities}
\label{asec8}
Here, we derive the vibration propagator in Eq. (\ref{Lmd1}) for two commonly used continuous spectral densities using contour integration methods. 
\subsection{Drude-Lorentz spectral density}
\begin{figure}
    \centering
    \includegraphics[width=0.5\linewidth]{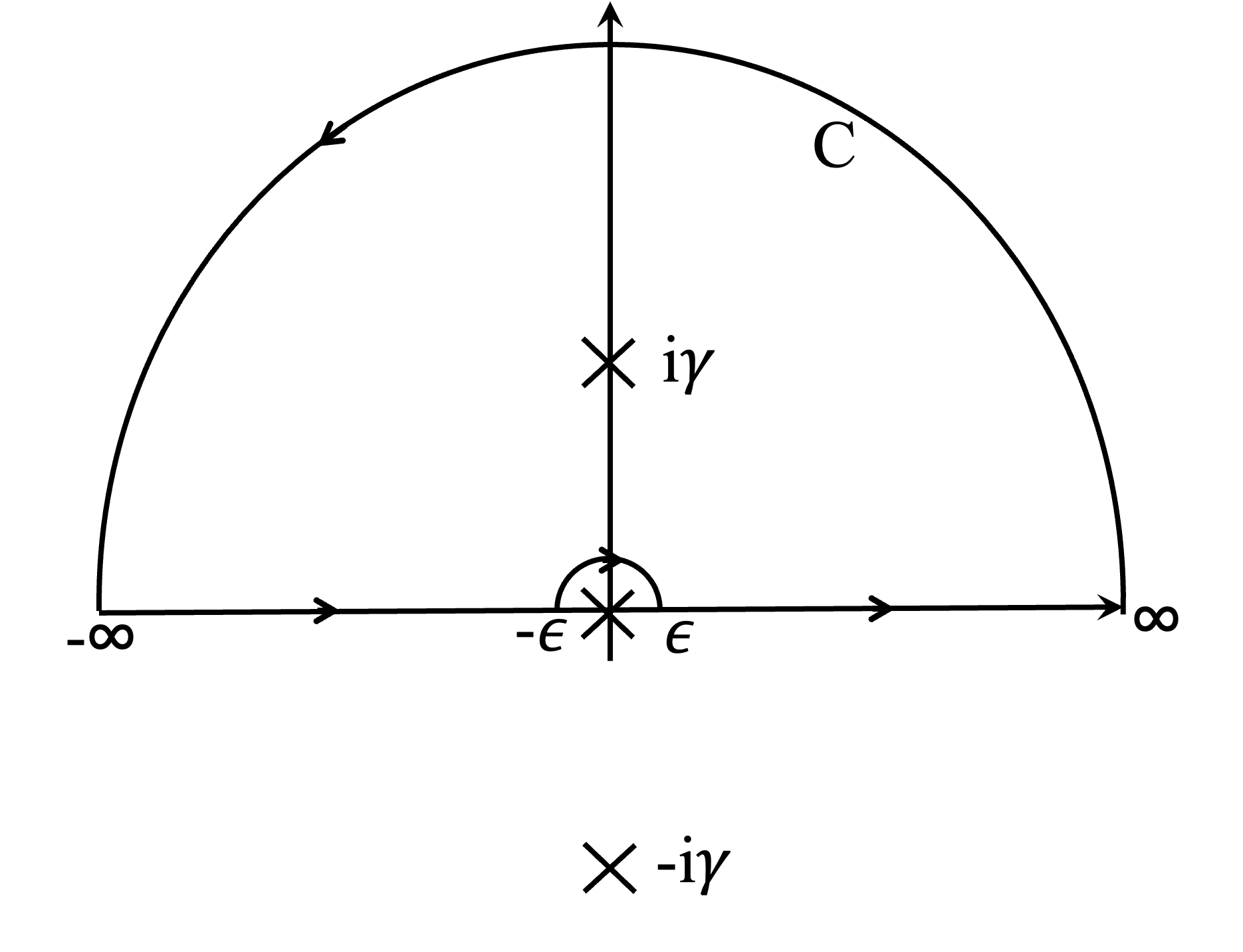}
    \caption{The contour choice for evaluating the vibration propagator for over-damped Drude-Lorentz spectral density. The cross signs denote the poles of the integrand.}
    \label{contour}
\end{figure}

For a Drude-Lorentz spectral density which is used for a complex polyatomic molecules \cite{ko2022dynamics} 
\begin{eqnarray}
\label{drude}
    J(\Omega) = \frac{2\lambda\Omega\gamma}{\pi(\Omega^2+\gamma^2)},
\end{eqnarray}
where $\gamma$ denotes the cut-off frequency, the vibration propagator integral can be expressed as the following complex integral:
\begin{eqnarray}
    \Lambda_1(t) = \frac{2\lambda\gamma}{\pi}\int_{0}^\infty \frac{\coth(\beta z/2)(\cos(tz)-1)+i\sin(tz)}{z(z^2+\gamma^2)} dz.
\end{eqnarray}
The integrand consists two poles of order 1 at $z=\pm i\gamma$ and a pole at $z=0$ of order 2 (since $\coth(\beta z/2)\approx 2/(\beta z)$ for $z\rightarrow 0$). We evaluate the following integrals: 
\begin{eqnarray}
\label{eq65}
    I_1(t) = \int_{-\infty}^\infty \frac{\coth(\beta z/2)(e^{itz}-1)}{z(z^2+\gamma^2)} dz \,\,\,\,\,\, \text{, and, }\,\,\,\,\,\, I_2(t) = \int_{-\infty}^\infty \frac{(e^{itz})}{z(z^2+\gamma^2)} dz.
\end{eqnarray}
From these two integrals, it is evident that the real-part $\Re(\Lambda_1(t))\equiv (\lambda\gamma/\pi)\Re(I_1(t))$ and the imaginary-part $\Im(\Lambda_1(t))\equiv (\lambda\gamma/\pi)\Im(I_2(t))$. Considering the poles of the integrand, we choose a semicircular arc contour in the positive imaginary hemisphere in the $z$ plane with an $\epsilon$-arc indentation at $z=0$ (see Fig.~\ref{contour}). 

Thus, for $t\geq0$ the integral can be represented as a contour integral and upon applying the residue-theorem,
\begin{eqnarray}
    \oint_{C} f(z) dz=\int_{-\infty}^{-\epsilon} f(z) dz + \int_{\epsilon-\text{arc}} f(z) dz +\int_{\epsilon}^\infty f(z) dz+ \int_{C}f(z) dz = 2\pi i \,\,\underset{z=i\gamma}{\text{Res }}[f(z)],
\end{eqnarray}
where $f(z)$ represents the integrand and $\text{Res}[.]$ represents the residue. The integral along the bigger semicircle $C$ vanishes due to Jordan's lemma. Thus, in the limit $\epsilon\rightarrow 0$, 
\begin{eqnarray}
    \int_{-\infty}^{\infty} f(z) dz = 2\pi i \,\,\underset{z=i\gamma}{\text{Res }}[f(z)]-
    \lim_{\epsilon\rightarrow 0}
    \int_{\epsilon-\text{arc}} f(z) dz,
\end{eqnarray}
The residue at $z=i\gamma$ of the integrand of $I_1(t)$ is:
\begin{eqnarray}
    \underset{z=i\gamma}{\text{Res }}\left[ \frac{\coth(\beta z/2)(e^{itz}-1)}{z(z^2+\gamma^2)}\right] = \frac{\coth(i\beta \gamma/2)(e^{-\gamma t}-1)}{i\gamma(2i\gamma)} = \frac{\text{cot}(\beta \gamma/2)(1-e^{-\gamma t})}{2i\gamma^2}.
\end{eqnarray}
The integral along the $\epsilon-$arc is evaluated as follows,
\begin{eqnarray}
    \underset{\epsilon\rightarrow 0}{\lim}\int_{\epsilon-\text{arc}}  \frac{\coth(\beta z/2)(e^{itz}-1)}{z(z^2+\gamma^2)} dz = \underset{\epsilon\rightarrow 0}{\lim}\int_\pi^0 d\theta\,\, i\epsilon e^{i\theta}\frac{\coth(\beta \epsilon e^{i\theta}/2)(e^{it\epsilon e^{i\theta}}-1)}{\epsilon e^{i\theta}(\epsilon^2e^{2i\theta}+\gamma^2)} \\
    =\frac{i}{\gamma^2}\int_\pi^0 d\theta\,\, \underset{\epsilon\rightarrow 0}{\lim}\,\,\frac{(e^{it\epsilon e^{i\theta}}-1)}{\text{tanh}(\beta \epsilon e^{i\theta}/2)}= \frac{2t\pi}{\gamma^2\beta} \,\,\,\,\,\,\text{ (using L'Hôpital's rule)}
\end{eqnarray}
Thus, for $t\geq0$ , the integral becomes: 
\begin{eqnarray}
    I_1(t\geq0) = -\frac{2\pi t}{\gamma^2\beta} + \frac{\pi \cot(\beta\gamma/2)(1-e^{-\gamma t})}{\gamma^2}.
\end{eqnarray}
For $t<0$, choosing a similar contour in the lower hemisphere in the imaginary axis in Fig.~\ref{contour} obtains:
\begin{eqnarray}
    I_1(t<0) = \frac{2\pi t}{\gamma^2\beta} - \frac{\pi \text{cot}(\beta\gamma/2)(1-e^{\gamma t})}{\gamma^2}.
\end{eqnarray}
Similarly, we can evaluate $I_2(t)$:
\begin{eqnarray}
     I_2(t) = \begin{cases} 
      \frac{i\pi}{\gamma^2}(1-e^{-\gamma t}) & t\geq 0 \\
      -\frac{i\pi}{\gamma^2}(1-e^{\gamma t}) & t<0 \\
   \end{cases}
\end{eqnarray}

Thus, the total vibration propagator takes the following form:
\begin{equation}
\label{eqs79}
\begin{aligned}
    \Lambda_1(t) = -\frac{2\lambda}{\gamma}\left(\frac{2\vert t\vert}{\beta} -\text{cot}(\beta\gamma/2)\left(1-e^{-\gamma\vert t\vert}\right)\right)
    + i\frac{2\lambda}{\gamma}\text{sgn}(t)(1-e^{-\gamma\vert t\vert}).
\end{aligned}
\end{equation}

Equation~\eqref{eqs79}, together with Eqs.~\eqref{lm2} and~\eqref{scatterp} in the main text, is used to compute the scattered pulse for this Drude–Lorentz spectral density and the corresponding Fisher informations shown in Fig.~\ref{qfi-cont}(a,c).

\subsection{Brownian-oscillator spectral density}

\begin{figure}
    \centering
    \includegraphics[width=0.5\linewidth]{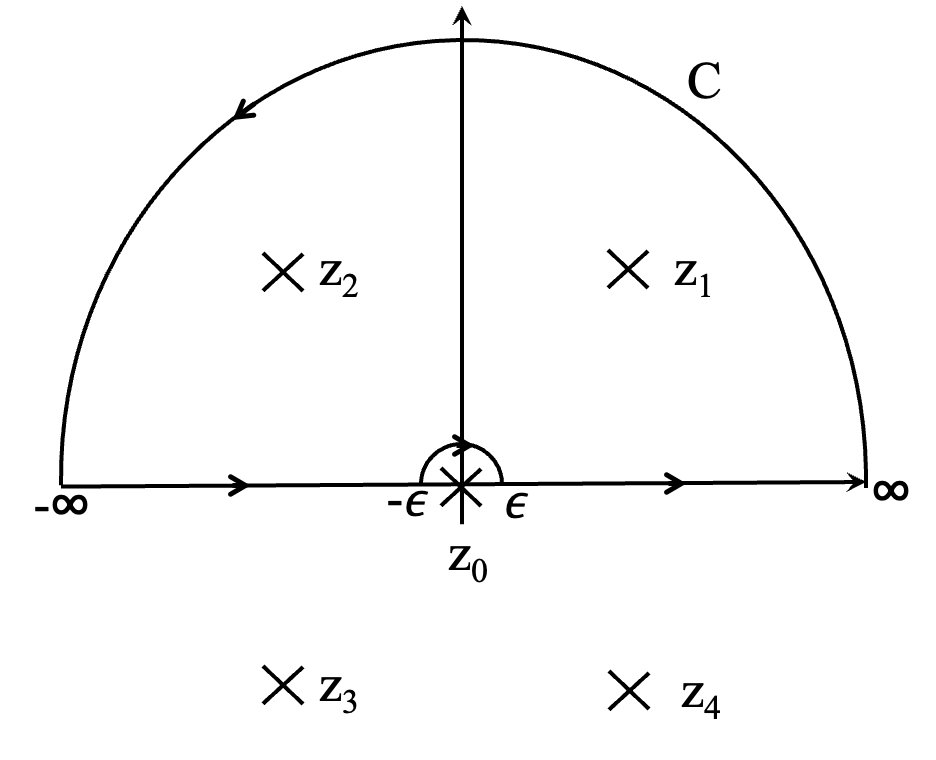}
    \caption{The contour choice for evaluating the vibration propagator for Brownian oscillator spectral density. The cross signs denote the poles of the integrand labeled by $z_k$ ($k=0,1,2,3,4$) which are given in the text.}
    \label{contour2}
\end{figure}

For Brownian-oscillator spectral density often used for quantum dot systems \cite{PhysRevLett.123.093601}, the spectral density is given by:
\begin{eqnarray}
\label{brown}
    J(\Omega) = \frac{2\lambda\Omega\gamma\Omega_0^2}{\pi((\Omega^2-\Omega_0^2)^2+\gamma^2\Omega^2)},
\end{eqnarray}
 where $\Omega_0$ is the vibrational characteristics frequency, the vibration propagator integral can be expressed as the following complex integral:
\begin{eqnarray}
    \Lambda_1(t) = \frac{2\lambda\gamma\Omega_0^2}{\pi}\int_{0}^\infty \frac{\coth(\beta z/2)(\cos(tz)-1)+i\sin(tz)}{z((z^2-\Omega_0^2)^2+\gamma^2z^2)} dz.
\end{eqnarray}
The integrand has 4 poles of order 1 at $z= \pm \Omega_c \pm i\gamma/2$ and a pole of order 2 at $z=0$. Here, $\Omega_c = \sqrt{\Omega_0^2-\gamma^2/4}$, where we assumed the under-damped condition: $\Omega_0>\gamma/2$. Let us denote the poles as follows: $z_0=0, z_1 = \Omega_c+ i\gamma/2, z_2 = -\Omega_c+ i\gamma/2,z_3 = -\Omega_c- i\gamma/2,z_1 = \Omega_c- i\gamma/2 $ based on their quadrant locations in Fig.~\ref{contour2}. 
Similar to Eq.~\eqref{eq65} we define the following integrals that will be evaluated,
\begin{eqnarray}
    I_1(t) = \int_{-\infty}^\infty \frac{\coth(\beta z/2)(e^{itz}-1)}{z((z^2-\Omega_0^2)^2+\gamma^2z^2)} dz \,\,\,\,\,\, \text{, and, }\,\,\,\,\,\, I_2(t) = \int_{-\infty}^\infty \frac{e^{itz}}{z((z^2-\Omega_0^2)^2+\gamma^2z^2)} dz.
\end{eqnarray}
From these two integrals, the real and imaginary part of the vibration propagator becomes $\Re(\Lambda_1(t))\equiv (\lambda\gamma\Omega_0^2/\pi)\Re(I_1(t))$ and  $\Im(\Lambda_1(t))\equiv (\lambda\gamma\Omega_0^2/\pi)\Im(I_2(t))$ respectively.
Now, choosing the contour in Fig.\ref{contour2} for $t\geq 0$, we can write the following,
\begin{eqnarray}
\label{eq78}
    \int_{-\infty}^{\infty} f(z) dz = 2\pi i \,\,\left(\underset{z=z_1}{\text{Res }}[f(z)]+\underset{z=z_2}{\text{Res }}[f(z)]\right)-
    \lim_{\epsilon\rightarrow 0}
    \int_{\epsilon-\text{arc}} f(z) dz.
\end{eqnarray}
The residue at $z_1$ is evaluated for $I_1$ as follows,
\begin{eqnarray}
\label{eq79}
    \underset{z=z_1}{\text{Res }}\left[\frac{\coth(\beta z/2)(e^{itz}-1)}{z((z^2-\Omega_0^2)^2+\gamma^2z^2)}\right]
    = \frac{\coth(\beta (\Omega_c + i\gamma/2)/2)(e^{it(\Omega_c+i\gamma/2)}-1)}{(\Omega_c+i\gamma/2)(i\gamma)(2\Omega_c+i\gamma)(2\Omega_c)}\nonumber \\
    = \frac{\big(\text{sinh}(\beta\Omega_c)+i\,\text{sin}(\beta\gamma/2)\big)\big(e^{-\gamma t/2}e^{i\Omega_c t}-1\big)(\Omega_c-i\gamma/2)^2}{4i\gamma \Omega_c\Omega_0^4 \,\,\big(\text{cosh}(\beta\Omega_c)-\text{cos}(\beta\gamma/2)\big)},
\end{eqnarray}
and, for $I_2,$
\begin{eqnarray}
\label{eq80}
    \underset{z=z_1}{\text{Res }}\left[\frac{e^{itz}}{z((z^2-\Omega_0^2)^2+\gamma^2z^2)}\right]
    = \frac{e^{it(\Omega_c+i\gamma/2)}}{(\Omega_c+i\gamma/2)(i\gamma)(2\Omega_c+i\gamma)(2\Omega_c)}= \frac{e^{-\gamma t/2}e^{i\Omega_c t}(\Omega_c-i\gamma/2)^2}{4i\gamma\Omega_c\Omega_0^2}.
\end{eqnarray}
Similarly, the residues at $z=z_2$ for $I_1$ is calculated as follows,
\begin{eqnarray}
\label{eq81}
    \underset{z=z_2}{\text{Res }}\left[\frac{\coth(\beta z/2)(e^{itz}-1)}{z((z^2-\Omega_0^2)^2+\gamma^2z^2)}\right]
    = \frac{\coth(\beta (-\Omega_c + i\gamma/2)/2)(e^{it(-\Omega_c+i\gamma/2)}-1)}{(-\Omega_c+i\gamma/2)(i\gamma)(-2\Omega_c+i\gamma)(-2\Omega_c)} \nonumber \\
    = \frac{\big(-\text{sinh}(\beta\Omega_c)+i\,\text{sin}(\beta\gamma/2)\big)\big(e^{-\gamma t/2}e^{-i\Omega_c t}-1\big)(\Omega_c+i\gamma/2)^2}{-4i\gamma \Omega_c\Omega_0^4\,\,\big(\text{cosh}(\beta\Omega_c)-\text{cos}(\beta\gamma/2)\big)},
\end{eqnarray}
and, for $I_2$,
\begin{eqnarray}
\label{eq82}
    \underset{z=z_2}{\text{Res }}\left[\frac{e^{itz}}{z((z^2-\Omega_0^2)^2+\gamma^2z^2)}\right]
    = \frac{e^{it(-\Omega_c+i\gamma/2)}}{(-\Omega_c+i\gamma/2)(i\gamma)(-2\Omega_c+i\gamma)(-2\Omega_c)} 
    = \frac{e^{-\gamma t/2}e^{-i\Omega_c t}(\Omega_c+i\gamma/2)^2}{-4i\gamma \Omega_c\Omega_0^4}.
\end{eqnarray}
The integral along the $\epsilon-$ arc is evaluated for $I_1$,
\begin{eqnarray}
\label{eq83}
    \underset{\epsilon\rightarrow 0}{\lim}\int_{\epsilon-\text{arc}} dz \frac{\coth(\beta z/2)(e^{itz}-1)}{z((z^2-\Omega_0^2)^2+\gamma^2z^2)} = \underset{\epsilon\rightarrow 0}{\lim}\int_{\pi}^0 d\theta\,\, i\epsilon e^{i\theta} \frac{\coth(\beta \epsilon e^{i\theta}/2)(e^{itz}-1)}{\epsilon e^{i\theta}((\epsilon^2 e^{2i\theta}-\Omega_0^2)^2+\gamma^2\epsilon^2 e^{2i\theta})} = \frac{\pi t}{\beta\Omega_0^4},
\end{eqnarray}
and, for $I_2$,
\begin{eqnarray}
\label{eq84}
    \underset{\epsilon\rightarrow 0}{\lim}\int_{\epsilon-\text{arc}} dz \frac{e^{itz}}{z((z^2-\Omega_0^2)^2+\gamma^2z^2)} = \underset{\epsilon\rightarrow 0}{\lim}\int_{\pi}^0 d\theta\,\, i\epsilon e^{i\theta} \frac{e^{it\epsilon e^{i\theta}}}{\epsilon e^{i\theta}((\epsilon^2 e^{2i\theta}-\Omega_0^2)^2+\gamma^2\epsilon^2 e^{2i\theta})} = \frac{i\pi }{\Omega_0^4}.
\end{eqnarray}
In both cases we used L'Hôpital's rule. Putting Eq.~\eqref{eq79}-\eqref{eq84} in Eq.~\eqref{eq78} we evaluate $I_1(t)$ and $I_2(t)$ that leads to the real and the imaginary parts of the 
propagator which are given by:
\begin{equation}
\label{s90}
\begin{aligned}
    \Re(\Lambda_1(t)) = -\frac{2 \lambda\gamma\vert t\vert}{\beta\Omega_0^2}-\frac{\lambda}{\Omega_0^2\,\Omega_c (\cosh(\beta\Omega_c)-\cos(\beta\gamma/2))} \times
    \Bigg[\sinh(\beta\Omega_c)\big((\Omega_0^2-\gamma^2/2)\left(1-e^{-\frac{\gamma \vert t\vert}{2}}\cos(\Omega_c t)\right)\\
    -\gamma\Omega_c\,\text{sgn}(t)e^{-\frac{\gamma \vert t\vert}{2}}\sin(\Omega_c t)\big)
    - \sin(\beta\gamma/2)\left(\gamma\Omega_c\left(1-e^{-\frac{\gamma \vert t\vert}{2}}\cos(\Omega_c t)\right)+(\Omega_0^2-\gamma^2/2)\,\text{sgn}(t)e^{-\frac{\gamma \vert t\vert}{2}}\sin(\Omega_c t)\right)
    \Bigg],
\end{aligned}
\end{equation}
and,
\begin{equation}
\label{s91}
\begin{aligned}
    \Im(\Lambda_1(t)) = \frac{\lambda\gamma\,\,\text{sgn}(t)}{\Omega_0^2}+ \frac{\lambda e^{-\frac{\gamma \vert t\vert}{2}}}{\Omega_0^2\,\Omega_c}
    \Bigg[(\Omega_0^2-\gamma^2/2)\sin(\Omega_c t) - \text{sgn}(t)\gamma\Omega_c\, \cos(\Omega_c t)\Bigg].
\end{aligned}
\end{equation}

Equations~\eqref{s90} and~\eqref{s91}, together with Eqs.~\eqref{lm2} and~\eqref{scatterp} in the main text, are used to compute the scattered pulse for this Brownian oscillator spectral density, as well as the corresponding Fisher informations shown in Fig.~\ref{qfi-cont}(b,d).

\section{Fisher information for continuous vibrational spectral densities}
\label{asec9}
\begin{figure}
    \centering
    \includegraphics[width=0.8\linewidth]{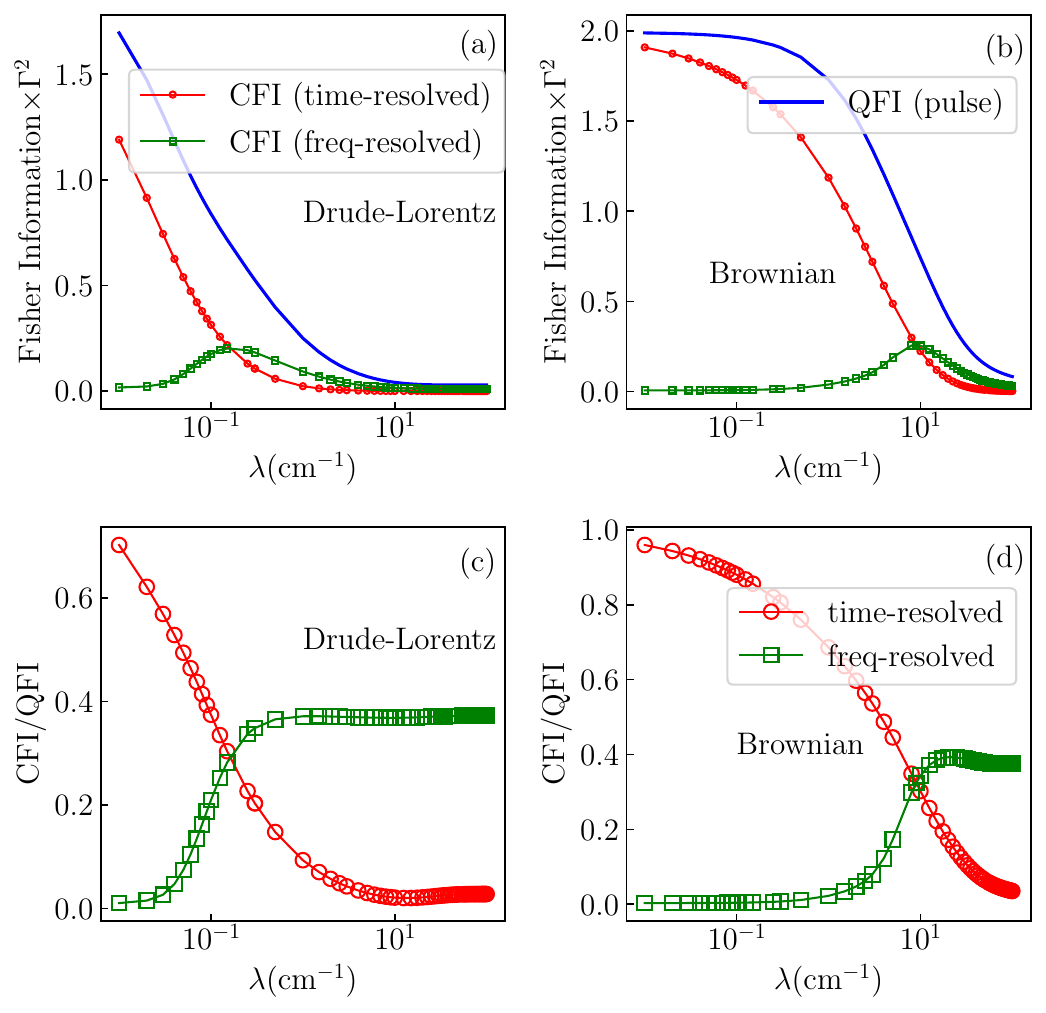}
    \caption{Fisher information (in units of $\Gamma^{2}$) as a function of the reorganization energy $\lambda$ for estimating the linewidth $\Gamma$ from the scattered pulse $\rho_\text{P}(\infty)$ when the vibrational environment has a continuum of frequency modes. Panels (a,b) show the quantum Fisher information (QFI, solid) and the classical Fisher information (CFI) obtained from time-resolved (dashed) and frequency-resolved (dot-dashed) photon-counting measurements. Panels (c,d) display the corresponding CFI/QFI ratios. Results in (a,c) correspond to a Drude--Lorentz spectral density, while (b,d) correspond to a Brownian-oscillator spectral density. The parameters used are: $\gamma = 10\,\mathrm{cm}^{-1}$, temperature $= 300\,\mathrm{K}$, an exponentially decaying incident pulse $\xi(t)$ with duration $T_{\sigma} = 1/\Gamma$, peak vibrational frequency $\Omega_{0} = 100\,\mathrm{cm}^{-1}$ (for Brownian oscillator in Eq.~\eqref{brown}), and $\Gamma = 0.15\,\mathrm{THz}$.}
    \label{qfi-cont}
\end{figure}

 In this section, we present the Fisher informations for estimating $\Gamma$ when the TLE is coupled to a continuum of vibrational modes. We characterize the vibrational coupling strength using the reorganization energy~$\lambda$, rather than the Huang--Rhys factor~$\lambda_0$ since the latter appropriately describes coupling contribution of a specific vibrational mode, whereas for a continuum of modes, the overall coupling is conventionally quantified by the reorganization energy
\begin{equation}
    \lambda = \int_0^\infty d\Omega\, \frac{J(\Omega)}{\Omega}.
\end{equation}
For a single mode of frequency $\Omega_0$, the two definitions coincide via $\lambda = \lambda_0 \Omega_0$. Therefore, we express coupling strengths in terms of $\lambda$ and consider values up to $100\,\mathrm{cm}^{-1}$, ensuring consistency with the parameters used in the preceding parts of this Supplemental material.

Figures~\ref{qfi-cont}(b,d) show the QFI, CFI, and their ratio as functions of the reorganization energy~$\lambda$ for $\Gamma$-estimation when the vibrational environment is described by a Brownian oscillator spectral density [Eq.~\eqref{brown}]. 
Figures~\ref{qfi-cont}(a) and (c) present the corresponding results for the Drude--Lorentz spectral density [Eq.~\eqref{drude}].   Notably, the conclusion that frequency-resolved measurements outperform time-resolved measurements at higher coupling strengths remains robust also for continuum environments.

\end{document}